\documentclass[10pt,journal,compsoc]{IEEEtran}
\usepackage{amsmath,amssymb,amsfonts,amsthm}
\usepackage{algorithmic}
\usepackage{graphicx}
\usepackage{textcomp}
\usepackage{xcolor}
\usepackage{multirow}
\usepackage{enumitem}
\usepackage{booktabs}
\usepackage{makecell}
\usepackage{array}
\usepackage{bm}
\usepackage{subfigure}
\usepackage[ruled,linesnumbered]{algorithm2e}

\newtheorem{definition}{Definition}

\newtheorem{task}{Task}

\ifCLASSOPTIONcompsoc
  \usepackage[nocompress]{cite}
\else
  \usepackage{cite}
\fi
\ifCLASSINFOpdf
\else
\fi
\newcommand{\tabincell}[2]{\begin{tabular}{@{}#1@{}}#2\end{tabular}} 

\begin{document}
%
\title{
SOUP: Spatial-Temporal Demand Forecasting and Competitive Supply
}

\author{Bolong Zheng,
	Qi Hu,
	Lingfeng Ming,
	Jilin Hu,
	Lu Chen,
	Kai Zheng,
	\and
	Christian S. Jensen,~\IEEEmembership{Fellow,~IEEE}
	\IEEEcompsocitemizethanks{
		\IEEEcompsocthanksitem Bolong Zheng, Qi Hu, Lingfeng Ming are with Huazhong University of Science and Technology.
		\protect\\
		E-mail: \{bolongzheng, huqi11, lingfengming\}@hust.edu.cn
		
		\IEEEcompsocthanksitem Jilin Hu and Christian S. Jensen are with Aalborg University.
		\protect\\
		E-mail: \{hujilin, csj\}@cs.aau.dk
		
		\IEEEcompsocthanksitem Lu Chen is with Zhejiang University.
		\protect\\
		E-mail: \{luchen\}@zju.edu.cn
		
		\IEEEcompsocthanksitem Kai Zheng is with University of Electronic Science and Technology of China.
		\protect\\
		E-mail: \{zhengkai\}@uestc.edu.cn
	}
	\thanks{Manuscript received April 19, 2005; revised September 17, 2014.}
}
\markboth{Journal of \LaTeX\ Class Files,~Vol.~14, No.~8, August~2015}%
{Shell \MakeLowercase{\textit{et al.}}: Bare Demo of IEEEtran.cls for Computer Society Journals}
%



\IEEEtitleabstractindextext{%
\begin{abstract}

We consider a setting with an evolving set of requests for transportation from an origin to a destination before a deadline and a set of agents capable of servicing the requests. In this setting, an assignment authority is to assign agents to requests such that the average idle time of the agents is minimized. An example is the scheduling of taxis (agents) to meet incoming passenger requests for trips while ensuring that the taxis are empty as little as possible.
In this paper, we study the problem of spatial-temporal demand forecasting and competitive supply (SOUP).
We address the problem in two steps. First, we build a granular model that provides spatial-temporal predictions of requests. Specifically, we propose a Spatial-Temporal Graph Convolutional Sequential Learning (\textsf{ST-GCSL}) model that predicts the requests across locations and time slots. Second, we provide means of routing agents to request origins while avoiding competition among the agents. In particular, we develop a demand-aware route planning (\textsf{DROP}) algorithm that considers both the spatial-temporal predictions and the supply-demand state.
We report on extensive experiments with real-world data that offer insight into the performance of the solution and show that it is capable of outperforming the state-of-the-art proposals.

\end{abstract}

}

\maketitle

\IEEEdisplaynontitleabstractindextext

%
\IEEEpeerreviewmaketitle

\section{Introduction}\label{sec:introduction}

The near-ubiquitous deployment of smartphones has enabled transportation network companies such as Didi Chuxing \cite{didi}, Uber \cite{uber}, and Lyft \cite{lyft} to operate ride-hailing platforms that enable the servicing of transportation requests by means of fleets of drivers. In this setting, drivers accept requests and move to the origins of requests to complete the requests. Such platforms have reduced significantly the amounts of time drivers are idle and the amounts of time spent waiting for service by prospective passengers, thus improving the traffic efficiency of a city. In this setting, historical requests provide insight into the movement patterns of passengers and drivers, which is beneficial for many applications such as traffic demand prediction, supply and demand scheduling, and route planning.

We study the problem of spatial-temporal demand forecasting and competitive supply (SOUP), which consists of forecasting spatial-temporal service requests, as well as planning routes for agents to active requests in a manner that minimizes the average idle time of all agents. Besides drivers (agents) looking for passengers (requests), this kind of competitive assignment problem also occurs in other urban transportation settings, e.g., drivers looking for parking and drivers looking for electric charging stations.

Our focus is on a population of drivers servicing an evolving set of requests for transportation from an origin to a destination within a given time window. The drivers are often called taxis. Most existing proposals on crowdsourced taxis focus either on how to better match taxis with service requests to maximize global revenue \cite{DBLP:conf/kdd/XuLGZLNLBY18, DBLP:conf/cikm/JinZ0LGQJTWWWY19, DBLP:conf/kdd/ZhangHMWZFGY17}, or how to learn taxi and passenger movement patterns from trajectory data to guide route planning \cite{DBLP:journals/ior/BravermanDLY19, DBLP:conf/kdd/QuZLLX14, DBLP:journals/tkde/YuanZZX13, DBLP:journals/tase/MiaoHLSZMHHP16}. Once a taxi drops off a passenger and completes a request, no further instructions are provided to the taxi to reduce the time it is idle before servicing the next request. Rather, taxis may either stay stationary or may move towards regions expected high demand, which may lead to competition. We aim to develop a data-driven solution that assigns a route to a taxi as soon as the taxi becomes idle such that the average time taxis are idle is minimized.

Overall, we address two sub-problems:
\begin{enumerate}
	\item \textbf{Dynamic request patterns.} In order to help agents service new requests quickly, we need to know the request patterns across the road network of a city. We first choose carefully a spatial granularity for partitioning a road network and temporal granularity for partitioning time in order to achieve accurate predictions of requests. We then build a corresponding model that predicts future requests. 
	\item \textbf{Competition among agents.} If all agents tend to move towards hot regions to find new requests, they will compete if the supply-demand ratio is high, which causes the so-called ``herding'' effect. To eliminate this effect, we develop a route planning strategy that assigns agents to destinations with supply-demand balance.
\end{enumerate}

The framework we propose consists of an offline and an online components. The offline component comprises an end-to-end deep learning model, called spatial-temporal graph convolutional sequential learning (\textsf{ST-GCSL}), that is capable of predicting requests at different locations and times. The online component comprises a demand-aware route planning (\textsf{DROP}) algorithm that exploits both the available spatial-temporal information on requests and the supply-demand state to guide idle agents.

The major contributions are summarized as follows:
\begin{itemize}
	\item We design an effective partitioning method that enables purposeful request prediction across space and time.
	\item We propose \textsf{ST-GCSL} to accurately predict future requests based on the partitioning.
	\item We develop \textsf{DROP} to assign routes that takes into account both the available spatial-temporal request information and the supply-demand state.
	\item We report on experiments that suggest the proposed \textsf{ST-GCSL} and \textsf{DROP} outperform the baseline methods.
\end{itemize}

The rest of the paper is organized as follows. We detail the problem addressed in Section \ref{sec:preliminaries}. In Section \ref{sec:datamodel}, we present the multi-level partitioning and the \textsf{ST-GCSL} algorithm. Section \ref{sec:routeplanning} presents the \textsf{DROP} algorithm. The experimental study is covered in Section \ref{sec:experiment}. The related work is the topic of Section \ref{sec:related}. Finally, Section \ref{sec:conclusion} concludes the paper.
\section{Preliminaries}\label{sec:preliminaries}

We proceed to introduce the background settings and to formalize the SOUP problem. Frequently used notation is summarized in Table \ref{tb:notation}.

\begin{table}
	\caption{Summary of Notations}
	\label{tb:notation}
	\centering
	\begin{tabular}{lp{6cm}}
		\toprule
		\textbf{Notation} & \textbf{Definition}\\
		\midrule
		$G = (V,E,W)$ & A road network  \\
		$\mathcal{A} = \{ a_i\}$ & A set of mobile agents \\
		${\Omega}=\{\omega_j\}$  & A set of requests \\
		$\mathcal{I}_{i}=\{I_{ik}\}$ & The set of idle times of agent $a_i$ \\
		$R=\{r_i\}$ & A set of regions on road network\\
		$T = \{t_i\}$ & A set of time slots during a day \\
		$r^*$ & The search route \\
		\hline
		$\mathcal{G}= (R,A)$ & A region correlation graph \\
		$A \in R^{N*N}$ & The adjacency matrix of graph \\
		$C$ & The number of channels for network input \\
		${D}_{j}^{i}$ & The number of requests in $r_i$ in time slot $t_j$ \\
		$\mathbf{D}^{i}$ & The request sequence of region $r_i$\\
		$\mathbf{D}_{j}$ & The request vector at time slot $t_j$ \\
		$\hat{\mathbf{D}}_{t+1}$ & The predicted request vector \\
		$\mathbf{\Psi}_t$ & The context feature vector at time slot $t$ \\
		$Z,Z^{'}$ & The input and output of an STCM \\
		$c_f$ & The final context features\\
		$X^l,X^{l+1}$ & The input and output of the $l$-th ST-Gate Block \\
		$\boldsymbol{\Gamma}^l_{*\boldsymbol{\tau}}$ & The temporal convolution kernel at $l$-th layer \\
		$\boldsymbol{\Theta}^l_{\star\mathcal{G}}$ & The spectral kernel of GC at $l$-th layer \\
		\bottomrule
	\end{tabular}
\end{table}

\subsection{Settings}
The problem setting encompasses four types of entities: a road network, mobile agents (taxis), requests (passengers), and an assignment authority. 

\begin{definition}[Road Network]
A road network is defined as a weighted directed graph $G=(V,E,W)$, where $V$ is the set of nodes, $E$ is the set of edges, and $W$ is the set of edge weights. Each edge $e(u,v) \in E$ that starts from node $u$ to $v$ has a positive weight $w(u,v) \in W$, i.e., travel time on the edge.
\end{definition}

\begin{definition}[Mobile Agent]
We assume a population of mobile agents $\mathcal{A}=\{a_i\}$, which is introduced into the system at once at the beginning. Each $a_i$ has an original location $l_i$ on the road network and labeled as empty. The population of agents is fixed throughout the operation and has cardinality $|\mathcal{A}|$. 
\end{definition}


\begin{definition}[Request]
We assume an evolving set of requests ${\Omega}=\{\omega_j\}$ that are introduced into the system in a streaming fashion. Each request $\omega_j=(o,d,t^o,t^*)$ has an origin $o$, a destination $d$, an introduction time $t^o$, and a maximum life time (MLT) $t^*$. A request that is not serviced by $t^*$ time units after $t^0$ is automatically removed from the system, an outcome that we call request expiration.
\end{definition}

When a request enters the system, the agent that meets the following conditions is assigned to the request by the assignment authority:
\begin{enumerate}
	\item The agent is \textit{empty}.
	\item The agent is the agent that is \textit{nearest} to the request.
	\item The \textit{shortest-travel-time} from the agent to the request enables the agent to reach the request before it expires.
\end{enumerate}

Once an agent is assigned to a request, the agent is labeled as occupied, and the request is removed from the system. Then the agent moves to the request (for pick-up) and then to the request destination (for drop-off), both along the \textit{shortest-travel-time path} in the road network. 
When the agent arrives at the destination, it is labeled as empty. If no agent meets the above conditions, the request remains in the system until an agent meets the conditions or the request expires. 

Once an agent becomes empty, the assignment authority plans a so-called \textbf{search route} $r^*$ for it to serve potential requests. If the agent finishes traversing its search route without having been assigned a new request, it is assigned a new search route.

It is worth noting that an agent neither knows when and where requests will appear nor has any information about other agents. They are not allowed to communicate or collaborate with each other.

\begin{definition}[Idle Time]
The idle time of an agent is the amount of time from when the agent is labeled as empty to when it is assigned to a request. An agent may experience multiple idle times in a day, each corresponding to traveling on a search route. 
We denote $\mathcal{I}_{i}=\{I_{ik}\}$ as the set of idle times of agent $a_i$'s search routes, where $I_{ik}$ is the idle time of $a_i$'s $k$-th search route $r^*_{ik}$.
\end{definition}

In order to reduce the idle time when planning a search route for an idle agent, we build an accurate request data model that the assignment authority can use to make decisions.
The data model is a software module that is shared with all agents and that is used to represent request patterns and predict future requests. 

\subsection{Problem Statement}

The SOUP problem consists of two tasks: (1) spatial-temporal demand forecasting; and (2) competitive supply.

\begin{task}[Spatial-Temporal Demand Forecasting]
Given a historical set of requests {$\Omega$}, we aim to build a request data model to predict the numbers of requests at different locations and times.
\end{task}

\begin{task}[Competitive Supply]
Given a request data model, a set of agents $\mathcal{A}$ with original locations, and a stream of requests {$\Omega$}, we aim to plan search routes for taxis to serve potential requests and avoid the competition such that the average idle time of taxis in Eq. \ref{eq:averagest} is minimized.

\begin{equation}\label{eq:averagest}
\frac{1}{\sum_{a_i \in \mathcal{A}}|\mathcal{I}_i|} \sum_{a_i \in \mathcal{A}} \sum_{I_{ik} \in \mathcal{I}_i} I_{ik}
\end{equation}
\end{task}

\begin{figure}[tbp]
	\centering
	\includegraphics[width=.45\textwidth]{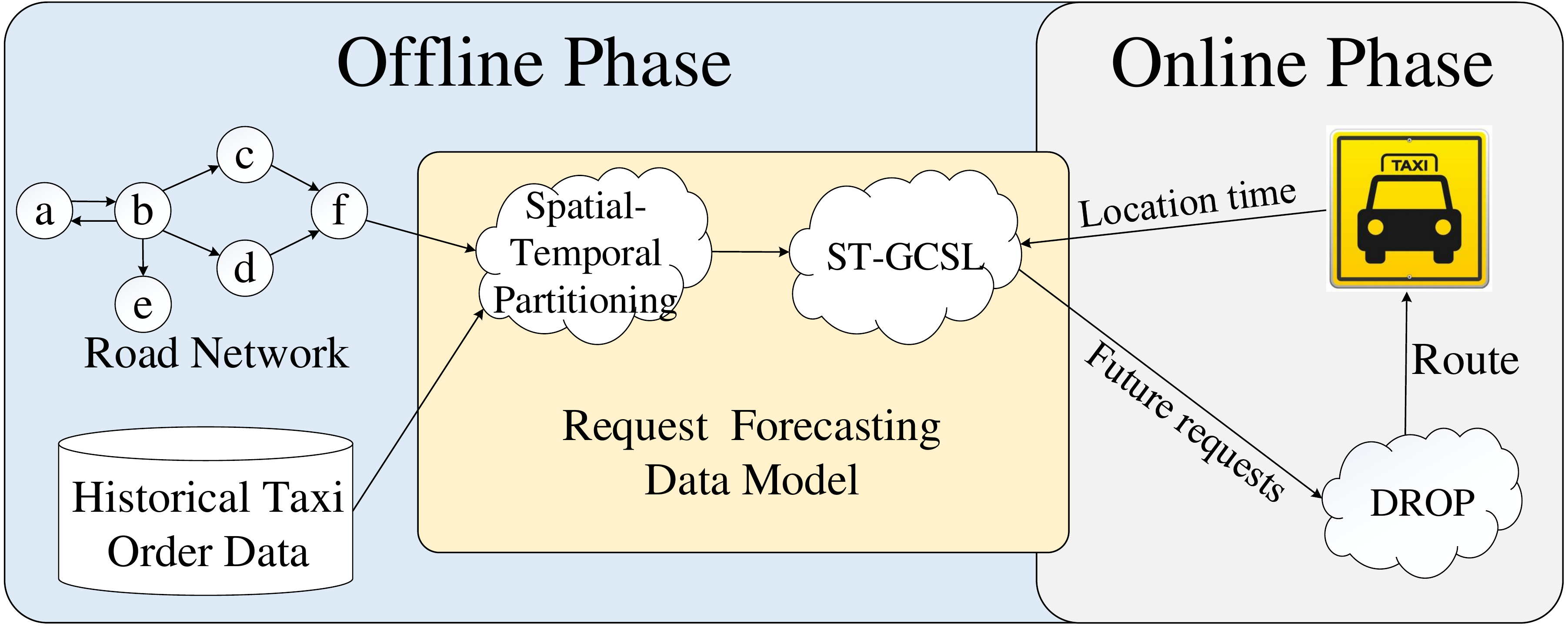}
	\caption{Framework Overview}
	\label{fig:framework}
\end{figure}

\subsection{Framework Overview}
To solve the above problem, we design a search route recommendation framework, as shown in Fig. \ref{fig:framework}. The processing pipeline includes offline and online components. For the former, we propose the \textsf{ST-GCSL} algorithm to train a request data model based on historical request data, i.e., taxi order data. The data model builds a partitioning on road network and predicts future requests for the partitions. For the latter component, we propose the \textsf{DROP} algorithm that computes a search route for an agent based on the location and time when it becomes idle and the supply-demand state in the near future. 

Once an agent is assigned to a request, it travels to the request for pickup and travels to the request's destination for dropoff. Afterwards, the assignment authority provides the agent with a search route. 
We limit our discussion to the forecasting and search processes, and we omit the details on the assignment of agents to requests.

\section{Spatial Temporal Request Forecasting} \label{sec:datamodel}

We proceed to introduce an end-to-end deep learning model, called spatial-temporal graph convolutional sequential learning (\textsf{ST-GCSL}), to predict where and when requests are likely to appear.
First, we introduce a spatial-temporal partitioning, based on which we build a region correlation graph. Then, we present the details of \textsf{ST-GCSL}. 
Unlike studies \cite{DBLP:conf/ijcai/WuPLJZ19,DBLP:conf/aaai/GengLWZYYL19,DBLP:conf/ijcai/BaiYK0S19} that ignore either spatial or short-term spatial-temporal dependencies, \textsf{ST-GCSL} captures all the spatial, temporal, short-term spatial-temporal dependencies, and context features to improve the prediction accuracy.

\subsection{Spatial-Temporal Partitioning}
\begin{figure}
	\begin{minipage}[htbp]{0.4\linewidth}
		\centering
		\subfigure[Hexagon Regions]{
		\includegraphics[width=\textwidth]{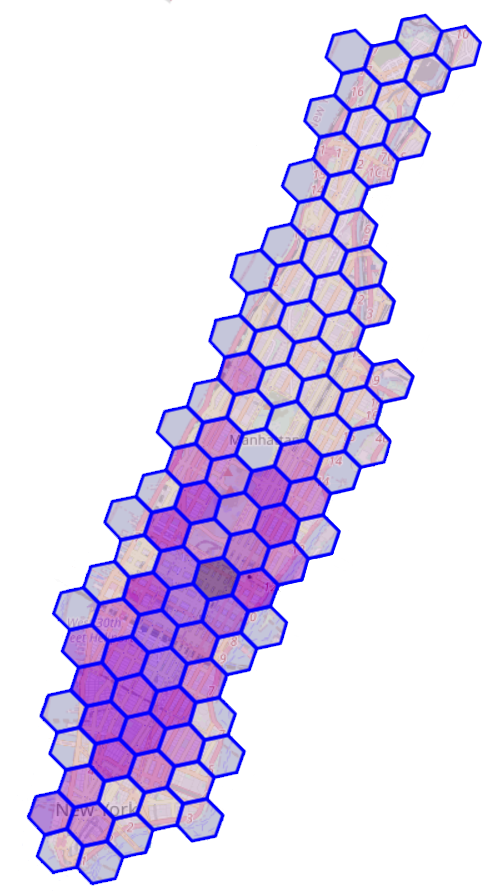}\label{fig:regions}
	}
	\end{minipage}
    \hfill
	\begin{minipage}[htbp]{0.6\linewidth}
		\centering
		\subfigure[Geo Neighbors]{
		\includegraphics[width=.5\textwidth]{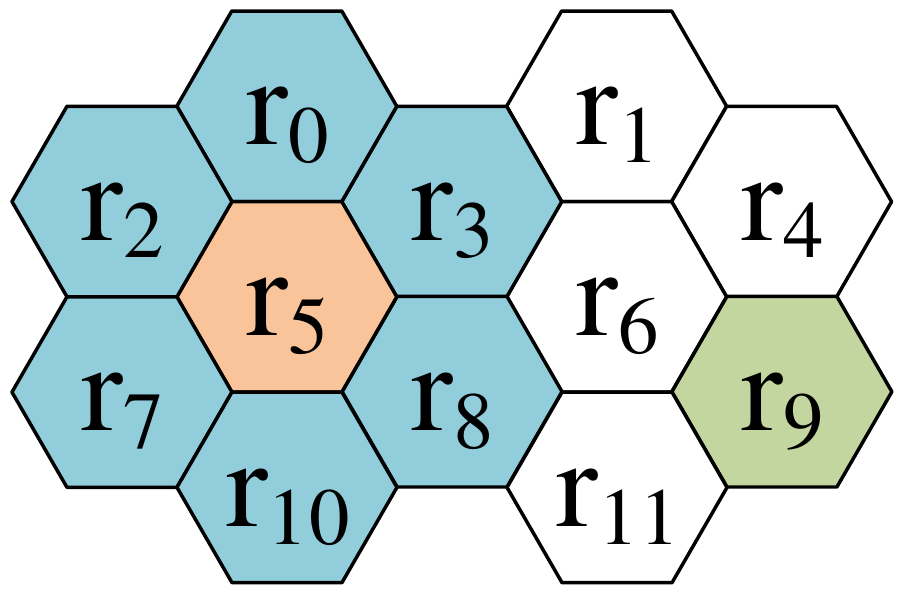}\label{fig:geo}}
		\vspace{-0.95cm}
	    \subfigure[Semantic Neighbors]{
		\includegraphics[width=.9\textwidth]{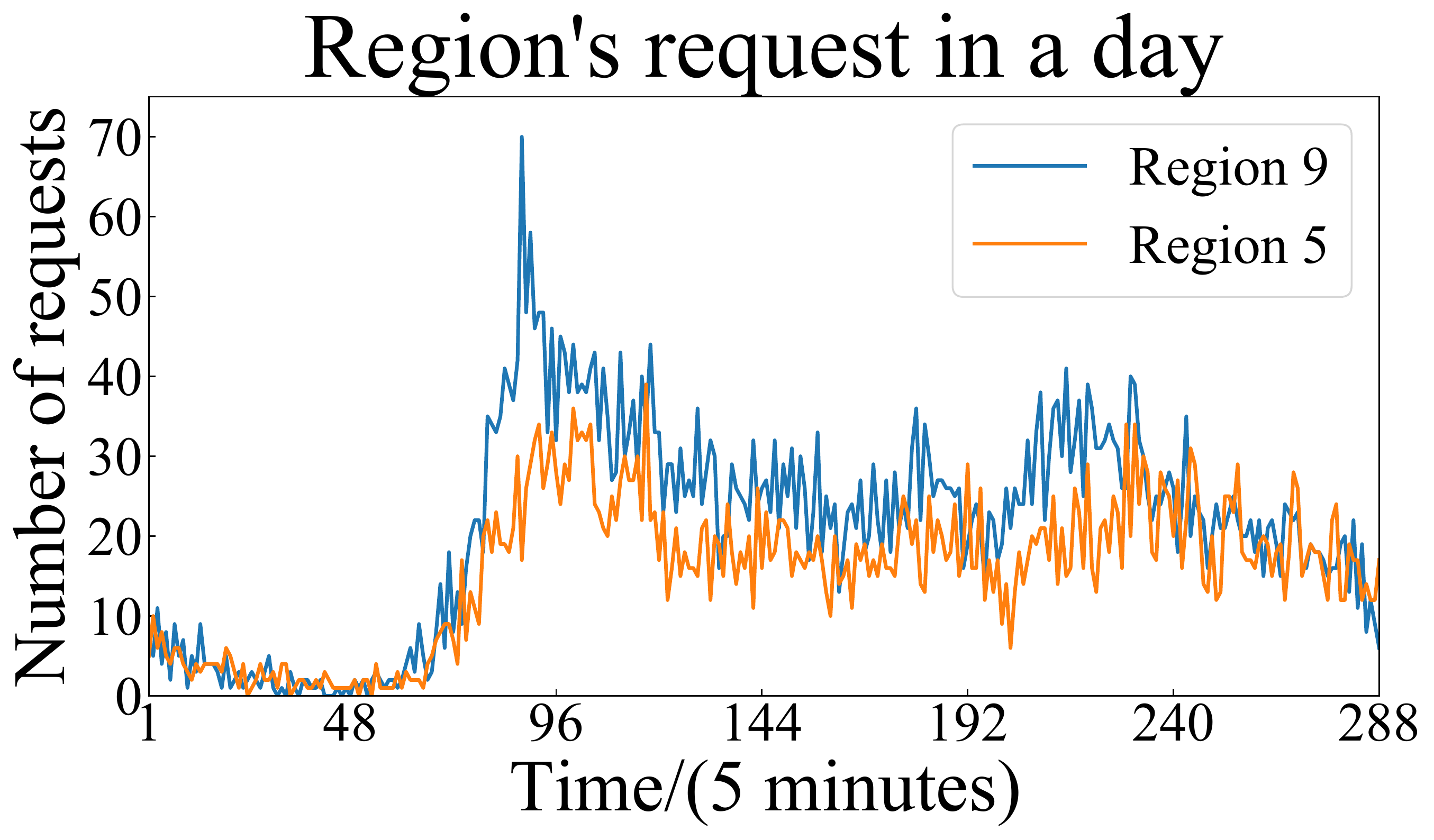}\label{fig:sem}
	}
	\end{minipage}\label{fig:neighbors}
    \caption{(a) The hexagon regions of Manhattan. Dark colors indicate high demands. (b) The geographical neighbors of $r_5$ are $r_0$, $r_2$, $r_3$, $r_7$, $r_8$, and $r_{10}$. (c) Since $r_9$ and $r_5$ have similar request patterns, they are semantic neighbors.}
	\label{fig:regions_and_neighbors}
\end{figure}

As shown in Fig. \ref{fig:regions}, we partition the road network into $N$ hexagons with Uber H3 library\footnote{https://github.com/uber/h3-java} and denote the set of regions by $R=\{r_1,r_2,...,r_N\}$. We partition a day into $M$ time slots and denote the set of time slots by $T=\{t_1,t_2,\dots,t_{M}\}$.
Let $D_{j}^{i} \in \mathbb{R}$ denote the number of requests in region $r_i$ in time slot $t_j$, we have:
\begin{equation}
D_{j}^{i} = |\{\omega|\ \omega.o \in r_i \land \omega.t^o \in t_j\}|.
\end{equation}
From the historical request dataset, we obtain a request sequence $ \langle \mathbf{D}_{1},\mathbf{D}_{2},\dots, \mathbf{D}_{{M}}\rangle$, where $\mathbf{D}_{j}=\{D_j^1,D_j^2,\dots,D_j^N\}$ denotes the numbers of requests of all the $N$ regions in time slot $t_j$, $1 \leq j \leq M$.

\subsection{Region Correlation Graph}\label{sec:region correlation graph}
Since the spatial dependency between regions can be captured accurately by a topology structure rather than by the Euclidean space \cite{DBLP:conf/ijcai/BaiYK0S19}, we transform the request prediction problem into a graph node prediction problem. Therefore, before we introduce the details of \textsf{ST-GCSL}, we first explain how to build a region correlation graph based on a historical request dataset.

We follow the existing study \cite{DBLP:conf/cikm/BaiYK0LY19} to derive a region correlation graph  $\mathcal{G} = (R,A)$, where the set of nodes $R$ is the region set introduced earlier, and $A$ is the edge set of $\mathcal{G}$ in the form of an adjacency matrix.

Specifically, as shown in Fig. \ref{fig:geo} and Fig. \ref{fig:sem}, to define $A$, we consider two types of region neighbors, i.e., \textbf{Geographical Neighbor} and \textbf{Semantic Neighbor}, based on whether two regions are geographically close or have similar request patterns. 
\begin{itemize}
	\item The geographical neighbor is based on the first law of geography \cite{doi:10.2307/143141}, ``near things are more related than distant things,'' which is used to extract spatial dependencies between a region and its adjacent regions. 
	\item The semantic neighbor is used to extract semantic correlations among regions with similar request patterns. The intuition is that distant regions may have similar request patterns if they have similar points of interest (POI). We adopt the Pearson Correlation Coefficient \cite{DBLP:conf/cikm/BaiYK0LY19} to quantify the request pattern similarity between regions. Let $\mathbf{D}^{i} = \{D_1^i,D_2^i,\dots,D_M^i\}$ represent the request sequence of region $r_i$ in the training data. The semantic similarity between $r_i$ and $r_{j}$ is defined as follows.
	\begin{equation}
	Sim(r_i,r_{j}) = Pearson(\mathbf{D}^{i},\mathbf{D}^{j})
	\end{equation}
	We consider regions $r_i$ and $r_{j}$ as semantic neighbors if $Sim(r_i,r_{j}) \geq \epsilon$, where $\epsilon$ is a threshold.
\end{itemize}

In the rest of the paper, we use neighbors to denote both geographical and semantic neighbors. The adjacency matrix $A$ is defined as follows:
\begin{equation}
\label{eq:adj}
A_{ij} = \begin{cases} 
1  & \mbox{if } r_i, r_{j} \mbox{ are neighbors} \\
0 & \mbox{otherwise}
\end{cases}
\end{equation}

\subsection{ST-GCSL Model}

\begin{figure}
	\centering
	\includegraphics[width=\linewidth]{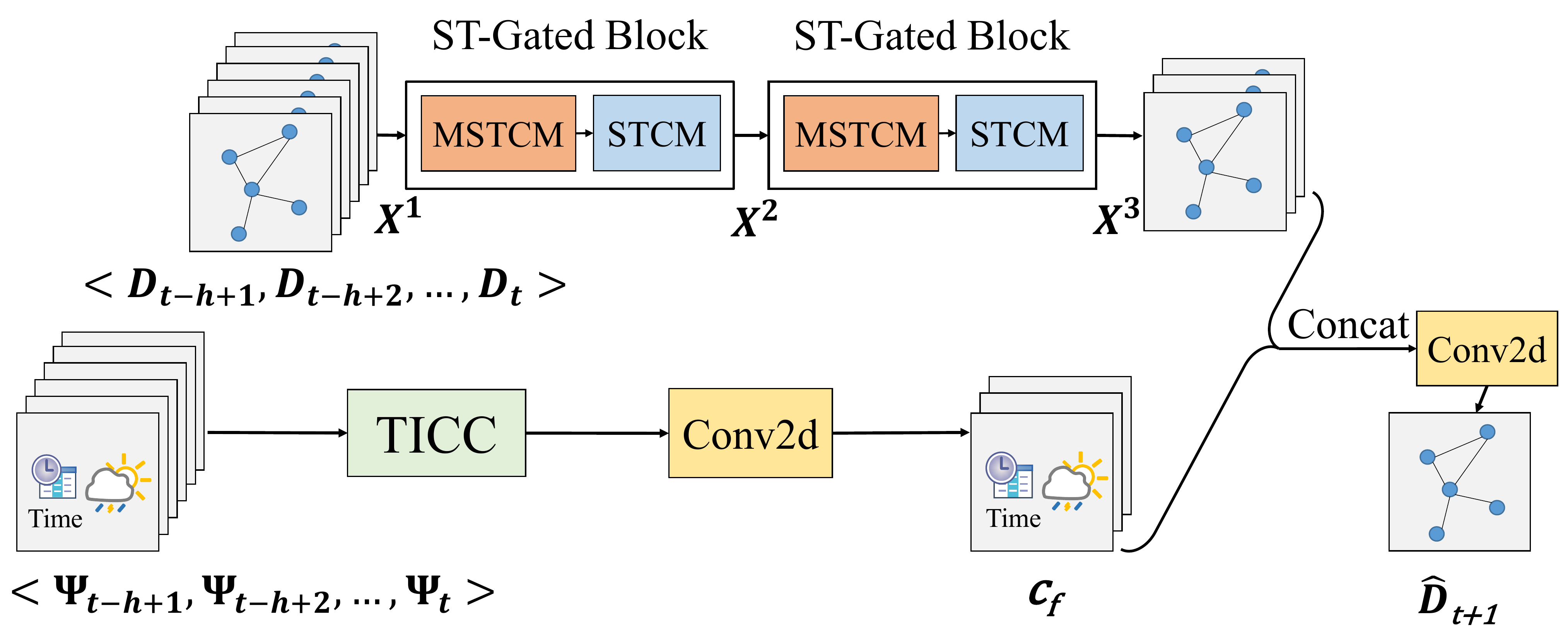}
	\caption{The architecture of \textsf{ST-GCSL} 
	}
	\label{fig:ST-GCSL}
\end{figure}

The structure of \textsf{ST-GCSL} is shown in Fig. \ref{fig:ST-GCSL}. Two parallel chains process the historical request sequence and the context feature sequence, respectively.
Consider the current time $t$.
\begin{enumerate}
\item In the upper chain, the input is the historical request sequence of the $h$ most recent time steps, $\langle \mathbf{D}_{t-h+1},\mathbf{D}_{t-h+2},\dots,\mathbf{D}_t\rangle$, which is processed by two Spatial-Temporal Gated Blocks (ST-Gated Blocks).
\item In the lower chain, the input is the corresponding context feature sequence,  $\langle \mathbf{\Psi}_{t-h+1},\mathbf{\Psi}_{t-h+2},\dots,\mathbf{\Psi}_t\rangle$, which is processed by Toeplitz Inverse Covariance-Based Clustering (TICC) \cite{DBLP:conf/ijcai/HallacVBL18} and 2D convolution (Conv2d). 
\end{enumerate}
The results of these two chains are concatenated and convoluted to return the predicted result at the next time step, i.e., $\mathbf{\hat{D}}_{t+1} \in \mathbb{R}^N$.

\subsubsection{Spatial-Temporal Gated Block (ST-Gated Block)}
The \textsf{ST-GCSL} model has two ST-Gated Blocks, each of which consists of two components: The Multiple Spatial-Temporal Convolutional Module (MSTCM) and The Spatial-Temporal Convolutional Module (STCM).
\begin{enumerate}
\item \textbf{MSTCM} is stacked by several STCMs (that we call inter-STCMs) and is used to capture short-term spatial-temporal dependencies.
\item \textbf{STCM} is used to capture long-term spatial-temporal dependencies. To distinguish it from the STCMs in MSTCM, we call it the outer-STCM.
\end{enumerate}

For the $l$-th ST-Gated Block, where $l \in \{1,2\}$, the input is the feature of the request sequence $X^l  \in \mathbb{R}^{h_l\times N\times C^l}$, where $h_l$ is the length of time steps, $N$ is the number of nodes (regions), and $C^{l}$ is the feature dimensionality.
When $l=1$, we have $C^l=1$.
The output is a new representation of the feature $X^{l+1} \in \mathbb{R}^{[h_l-2(k-1)]\times N\times C^{l+1}}$, where $k$ is the kernel size of convolution operation. Note that we use two ST-Gated Blocks, so the time step length $h$ decreases to $h-4(k-1)$.

\begin{figure*}
	\centering
	\subfigure[Spatial-Temporal Convolutional Module (STCM)]{\includegraphics[width=.48\linewidth]{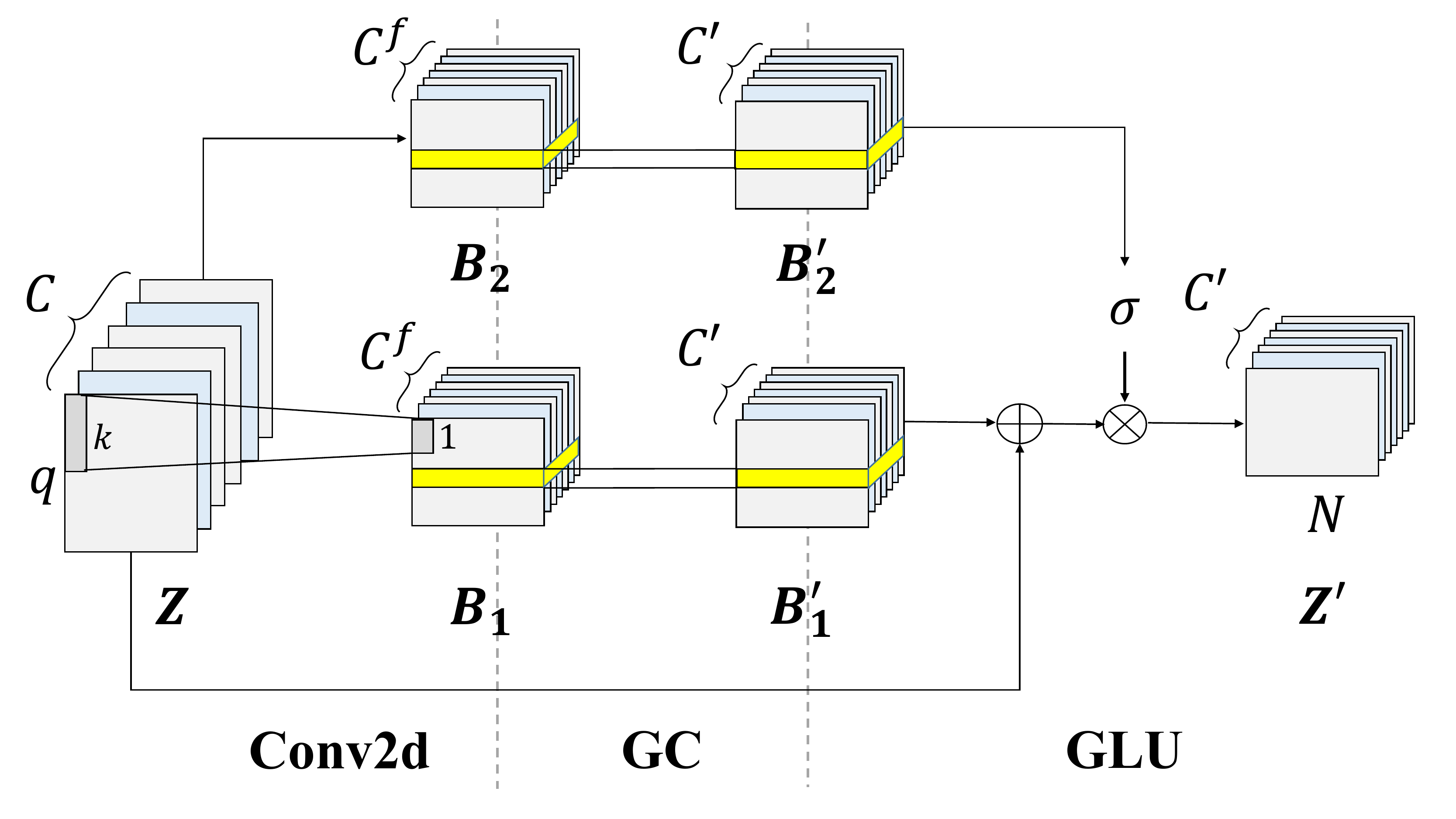}\label{fig:STCM}}
	\hfill
	\subfigure[Multiple Spatial-Temporal Convolutional Module (MSTCM)]{\includegraphics[width=.48\linewidth]{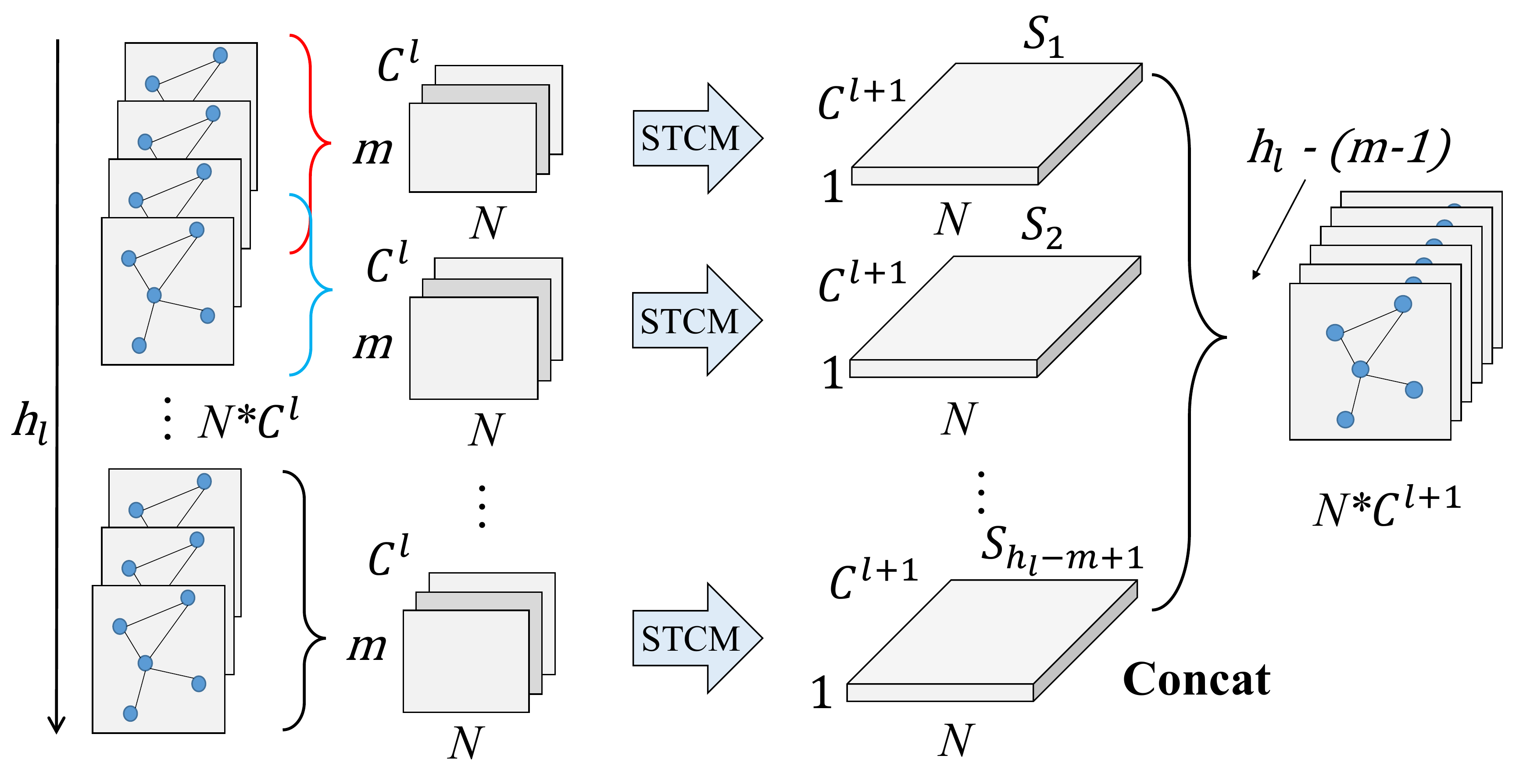}\label{fig:MSTCM}}
	\caption{Spatial-Temporal Gated Block
	}\label{fig:st_gate_block}
\end{figure*}

\textbf{STCM.} STCM is the key module of an ST-Gated Block. As shown in Fig. \ref{fig:STCM}, STCM consists of three operations: Conv2d, Graph Convolution (GC), and Gated Linear Unit (GLU) \cite{DBLP:conf/icml/DauphinFAG17}. 
The input of STCM is $Z\in \mathbb{R}^{q \times N \times C}$. Specifically, for outer-STCM, we have $q=h_l - (k-1)$. For inter-STCMs, we have $q=m$, where $m$ is the short-term time step length. 
The output is $Z^{'}\in \mathbb{R}^{[q-(k-1)] \times N \times C^{'}}$.

The input is first processed by Conv2d as follows.
\begin{equation}
\label{eq:cov2d}
\boldsymbol{\Gamma}_{*\boldsymbol{\tau}} * Z = B, 
\end{equation}
where $*$ denotes the convolution operator. $\boldsymbol{\Gamma}_{*\boldsymbol{\tau}}$ represents a total of $2\times C^f$ convolution filters, each of which has kernel size $k\times 1$, stride number 1, and no padding. The output of Conv2d is $B \in \mathbb{R}^{[q-(k-1)] \times N \times 2C^f}$, which is equally split into $B_1$ and $B_2$ along the feature dimension, i.e., $B_1, B_2 \in \mathbb{R}^{[q-(k-1)] \times N \times C^f}$. 

Then, $B_1$ and $B_2$ are fed into two GC layers separately, which is formulated as follows \cite{DBLP:conf/iclr/KipfW17}.
\begin{equation}
\label{GCN_without_GLU}
B_{\mu}^{'} = \boldsymbol{\Theta}_{\star\mathcal{G},\mu} \star B_{\mu} = \tilde{P}^{-\frac{1}{2}}\tilde{A}\tilde{P}^{-\frac{1}{2}}{B_{\mu}}{W_{\mu}},
\end{equation}
where $\mu \in \{1,2\}$, $\star$ denotes the graph convolution operator, and $\boldsymbol{\Theta}_{\star\mathcal{G},\mu}$ is a graph convolution filter. $\tilde{A} = I + A \in \mathbb{R}^{N \times N}$ is the adjacency matrix with self-looping. $\tilde{P}\in \mathbb{R}^{N \times N}$ is the degree matrix with $\tilde{P}_{ii} = \sum_{j}{\tilde{A}_{ij}}$ and $\tilde{P}_{ij} = 0$, $\forall i\neq j$. $B_{\mu}$ is the input and $W_{\mu}$ is the weight parameters.

Furthermore, we use a GLU to model the complex non-linearity in request forecasting, which is formulated as follows.
\begin{equation}
\label{GCN_with_GLU}
Z^{'} = \left[B_{1}^{'}  + Z \right]\otimes \sigma\left[B_{2}^{'}\right],
\end{equation}
where $\sigma$ is the Sigmoid function, and $\otimes$ denotes the Hadamard product. On the left half, a residual connection is utilized to avoid network degradation. The right half is a gate that regulates the information flow. Therefore, we have the output $Z^{'} \in \mathbb{R}^{[q-(k-1)] \times N \times C^{'}}$.

Generally, an STCM utilizes Conv2d to extract temporal dependencies and uses GC to capture spatial dependencies. Therefore, an STCM extracts short-term or long-term spatial-temporal dependencies based on the input time step size. Specifically, the inter-STCM is to extract short-term spatial-temporal dependencies with $q=m$, while the outer-STCM is to extract long-term spatial-temporal dependencies with $q=h_l-(k-1)$. Note that $m$ is smaller than $h_l-(k-1)$.

\textbf{MSTCM.}
As shown in Fig. \ref{fig:MSTCM}, MSTCM is stacked by multiple inter-STCMs along the time axis to capture the short-term spatial-temporal dependencies. For an MSTCM in the $l$-th ST-Gated Block, the input $X^l\in \mathbb{R}^{h_l\times N\times C^l}$ is split into $m$-Gram slices. Each slice $X_i^l\in \mathbb{R}^{m\times N\times C^l}$ is the input to the $i$-th inter-STCM, where $1\leq i \leq h_l-m+1$.
The output of MSTCM is $S=[S_1, S_2, \cdots, S_{h_{l}-m+1}]$, where $[\cdot,\cdot]$ is the concatenation operator, and $S_i \in \mathbb{R}^{[m-(k-1)]\times N\times C^{l+1}}$ is the output of the $i$-th inter-STCM.
According to Eqs.~\ref{eq:cov2d},~\ref{GCN_without_GLU}, and~\ref{GCN_with_GLU}, the operation of each inter-STCM is formulated as follows.
\begin{equation}
\label{eq:s_i}
S_i = [\boldsymbol{\Theta}^l_{\star\mathcal{G},1} \star (\boldsymbol{\Gamma}^l_{*\boldsymbol{\tau},1} *  X_i^l) + X_i^l] 
\otimes\sigma [\boldsymbol{\Theta}^l_{\star\mathcal{G},2} \star (\boldsymbol{\Gamma}^l_{*\boldsymbol{\tau},2} * X_i^l)].
\end{equation}
For the ease of representation, we use $m=k$ in all MSTCMs, so that $S_i \in \mathbb{R}^{1\times N\times C^{l+1}}$. 

To improve the processing efficiency, these inter-STCMs are executed in parallel such that MSTCM captures the short-term spatial-temporal dependencies of the entire input simultaneously. 

\subsubsection{Clustering Context Feature Sequence}
Due to the strong correlation between the request availability pattern and the extra features, e.g., weekdays, weekends, or rainy days, we introduce them as context features to improve the accuracy of request forecasting. As shown in Fig.~\ref{fig:ST-GCSL}, we employ clustering on the context features and determine which cluster the context features belong to. Then we add the cluster information as an additional feature.

The context feature $\mathbf{\Psi}_t$ at time step $t$ is a five-tuple (time of day, day of week, weather, holiday, events), which is clustered via TICC. Interested readers may refer to reference \cite{DBLP:conf/ijcai/HallacVBL18} for details. Thus, the context features of all $h$ time steps are encoded into a cluster label vector $c\in \mathbb{R}^{h \times 1}$. To concatenate with the output from the upper chain,  $c$ is processed by the following two steps on the spatial and temporal dimensions. First, $c$ is duplicated into a tensor $c_d\in \mathbb{R}^{h\times N\times 1}$ to expand the spatial dimension. Then, $c_d$ is processed by a convolution operation with a filter, whose kernel size is $[4(k-1)+1] \times 1$, so that the output is $c_f\in \mathbb{R}^{[h-4(k-1)]\times N \times 1}$.

\section{Demand-aware Route Planning}\label{sec:routeplanning}
We proceed to detail the demand-aware route planning (\textsf{DROP}) algorithm that guides idle agents based on the predicted request patterns such that the supply and demand are balanced across regions. Specifically, we first check whether the current time is during peak or non-peak hours. Then, we compute a weighted score for each candidate region that measures its degree of popularity. 
After that, we sample a destination region based on the weighted scores. 
Finally, we select a destination node in the destination region and provide the shortest-travel-time path to this node.

\subsection{Supply-Demand Analysis}
\begin{figure}
\centering
\subfigure[12:00, non-peak hours]{\includegraphics[width=.45\linewidth]{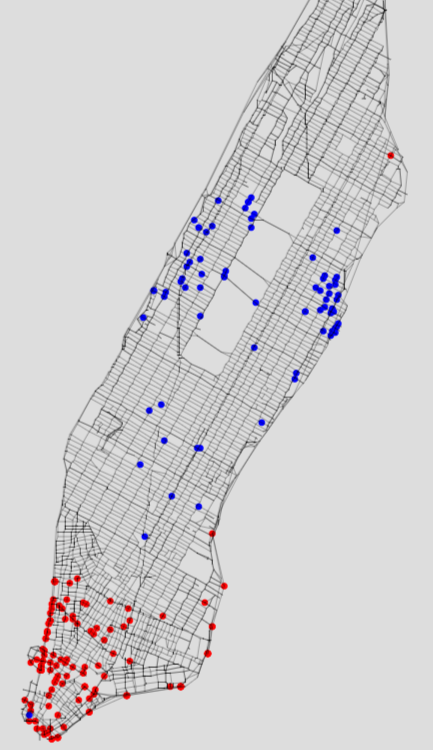}\label{fig:non-peak}}
\subfigure[20:00, peak hours]{\includegraphics[width=.45\linewidth]{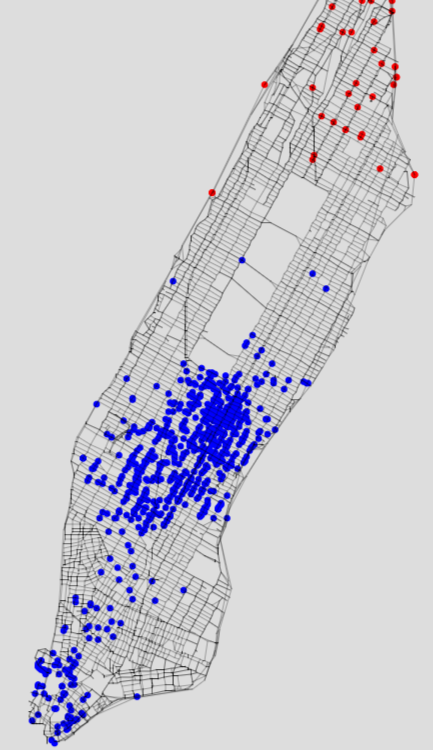}\label{fig:peak}}
\caption{Peak and non-peak hours (red dot: idle agent, blue dot: request)}\label{fig:agents_requests}
\end{figure}

To better understand the characteristics of peak and non-peak hours, we visualize the real-time states of requests and idle agents with the COMSET\footnote{https://github.com/JeroenSchols/COMSET-GISCUP} simulator, as shown in Fig. \ref{fig:agents_requests}. 
The New York TLC Trip Record YELLOW Data\footnote{https://www1.nyc.gov/site/tlc/about/tlc-trip-record-data.page} on June 1, 2016 is used as the simulation data, and we visualize two moments that correspond to peak and non-peak hours.

In Fig. \ref{fig:non-peak}, we observe that at 12:00 (non-peak hour), the number of requests is small, while most agents locate in the downtown area. Therefore, we need to spread agents across the network as much as possible to eliminate the ``herding effect" such that most requests can be served. Fig. \ref{fig:peak} shows that the number of requests exceeds by far the number of idle agents at 20:00 (peak hour), meaning that the demand by far exceeds the supply, and the request density in the midtown area is significantly higher than those in other areas. Therefore, to reduce the agent idle time, we need to guide all agents to the midtown area, rather scatter them across the road network.

We need to develop different routing strategies for the above two supply-demand states in peak versus non-peak hours. Therefore, we first check whether the current time is during peak or non-peak hours. 
Intuitively, the number of idle agents during peak hours is small.
Therefore, we compute a real-time ratio of the number of idle agents over the agent cardinality. If the ratio is smaller than a threshold $\delta$, the current moment is during peak hours. Otherwise, it is during non-peak hours.

\subsection{Computing Region Weighted Scores}
We utilize the results predicted by \textsf{ST-GCSL} to compute a weighted score for each region. Specifically, we first evaluate the popularity of each region based on its numbers of request origins and destinations, and we then compute region weighted scores for the candidate regions.

\textbf{Region Popularity Evaluation.}
Intuitively, the more request origins that locate in a region, the more popular the region is.
During peak hours, we only need to send agents to popular regions to serve requests. However, during non-peak hours, we have to consider the competition among agents. If a region is the destination of many requests, newly available agents will soon appear in that region. As a result, part of the requests in this region can be served quickly by these agents \cite{DBLP:conf/gis/HuMTZ19}. Therefore, the popularity of $r_i$ during $t_j$ is computed as follows.
\begin{equation}
\mathcal{P}_{i,j} = \begin{cases} 
o_{i,j}  & \mbox{during peak hours} \\
o_{i,j} - \lambda \times d_{i,j} & \mbox{during non-peak hours,}
\end{cases}
\end{equation}
where $o_{i,j}$ and $d_{i,j}$ represent the predicted number of request origins and destinations in $r_i$ during time slot $t_j$, respectively, and $\lambda$ balances the positive effect of request origins and the negative effect of request destinations. 

\textbf{Region Weighted Score.}
Since an agent may spend several time slots moving to the destination, the popularity of the destination region is not consistent in this process. Thus, we use the average popularity of the region in this process as the region weighted score. Furthermore, agents prefer to serve nearby requests rather than distant requests \cite{DBLP:conf/gis/KimPZ19}, so we use a discount factor to attenuate the region weighted score. For a region $r_i$ and time slot $t_j$, the region weighted score $\mathcal{W}_{i,j}$ is computed as follows.
\begin{equation}
    \mathcal{W}_{i,j} = \frac{\sum_{t=j}^{j+\Delta}\gamma^{t-j} \times \mathcal{P}_{i,j}}{\Delta},
\end{equation}
where $\Delta$ is the number of estimated time slots it takes an agent to move to the center of $r_i$, and $\gamma$ is the discount factor.

In addition, agents can be assigned to serve requests not only in the current region, but also in nearby regions. We conduct a spatial expansion \cite{DBLP:conf/icdm/WangQTYZ18} by taking the scores of nearby regions into consideration. With $R_1(r_i)$ representing the first-order neighbors of $r_i$, a new region weighted score $\mathcal{W}_{i,j}^*$ is computed as follows.
\begin{equation}
\label{eq:score}
    \mathcal{W}_{i,j}^* = \alpha \times \mathcal{W}_{i,j} + (1-\alpha) \times \frac{\sum_{r_k\in R_1(r_i)}\mathcal{W}_{k,j}}{|R_1(r_i)|}
\end{equation}
The first term in Eq. \ref{eq:score} is the previous score of $r_i$, and the second term is the average score of $r_i$'s first-order neighbors, and parameter $\alpha$ is used to balance their weights.

\subsection{Route Planning}
The route planning algorithm \textsf{DROP} computes a search route for an agent when the agent becomes idle. 
It is worth noting that the search strategy is state-dependent, i.e., the search strategies for peak and non-peak hours are different.

\textbf{Candidate Regions Generation.} To limit the length of the search route and to reduce the search space, we only consider the $L$-order neighbors \cite{DBLP:conf/icdm/WangQTYZ18} of the agent. For the agent's current region $r$, let $R_L(r)$ represent a set of regions, where each region is the $L$-order neighbors of $r$. Note that $R_0(r)=\{r\}$. We add all regions from $R_0(r)$ to $R_L(r)$ to form the candidate regions $R^*=\cup_{i=0}^{L}R_i(r)$. 

\textbf{Destination Region Determination.} In order to dispatch agents according to the request distribution, we first compute a weighted score for each region in $R^*$. For peak hours, as shown in Fig. \ref{fig:peak}, the demand exceeds by far the supply, and we only keep the $n$ regions with the highest weighted scores in $R^*$ and remove the remaining regions from $R^*$. For non-peak hours, we keep $R^*$ unchanged. Then, we employ a weighted sampling algorithm \cite{DBLP:conf/gis/HuMTZ19} to sample a destination region from $R^*$. For a candidate region $r_i$, the probability of $r_i$ being sampled is:
\begin{equation}
	\Pr(r_i)=\frac{\mathcal{W}_{i,j}^*}{\sum_{r_k \in R^*}\mathcal{W}_{k,j}^*}
\end{equation}

\textbf{Planning Search Route.} To determine the destination, we first select the top 15\% nodes with most requests in the destination region as potential destinations. Then, we perform weighted sampling again to find a node as the destination. Finally, a search route to this node along the shortest-travel-time path is assigned to the agent. Note that the destination selection improves the performance to a limited degree. Instead of training the \textsf{ST-GCSL} on the nodes of the road network, we simply use the average number of requests in the historical dataset as the weighted score of each node. Moreover, the 15\% of nodes in each region at each time slot can be precomputed to reduce the time cost of route planning.

\section{Experimental Study}\label{sec:experiment}
The \textsf{ST-GCSL} model is implemented in Python3 with Tensorflow, and the \textsf{DROP} algorithm is implemented in Java. The experiments are run on a Windows machine with an Intel 2.8GHz CPU and 12GB memory.

\subsection{Experimental Settings}
\noindent
\textbf{Dataset Description.} We use two real datasets, called New York and Haikou\footnote{https://outreach.didichuxing.com/app-vue/HaiKou?id=999}, to evaluate our method. Table \ref{Tab:datasets} summarizes statistics of the datasets. A request corresponds to a trip record. 

\begin{itemize}
	\item \textbf{New York:} The New York TLC Trip Record YELLOW Data includes the records with both pick-up and drop-off information of Yellow Cab taxis in New York City. We extract the data from January through June 2016 for evaluation. The weather data is extracted from New York Central Park\footnote{http://www.meteomanz.com/index?l=1\&cou=4030\&ind=72506}.
	\item \textbf{Haikou:} The Haikou dataset contains taxi order data from Haikou city for the period from May to October 2017, including the coordinates of origins and destinations, as well as the order type, the travel category, and the number of passengers. The weather data\footnote{https://data.cma.cn/} is extracted as the context features.
\end{itemize}

\begin{table}
	\caption{Dataset Statistics}
	\label{Tab:datasets}
	\centering
	\begin{tabular}{ccccc}
		\toprule
		\textbf{Dataset} &\textbf{\# of Orders} &\textbf{\# Edges}&\textbf{\# Nodes} & \textbf{Size (GB)}\\
		\midrule
		New York & 69,406,526 & 9,542 &4,360 &10.1\\
		Haikou & 12,374,094 & 8,034 & 3,298 & 2.28\\
		\bottomrule
	\end{tabular}
\end{table}

\textbf{Methods for Comparison.} For the request forecasting, we compare \textsf{ST-GCSL} with the following methods:
\begin{itemize}
	\item \textbf{HA}: A historical average model that treats the average number of requests as the predicted value.
	\item \textbf{VAR}\cite{hamilton1994time}: A Vector Auto-Regression model that is used to analyze multivariate time series data. 
	\item \textbf{LSTM}\cite{DBLP:journals/neco/HochreiterS97}: A Long Short-Term Memory Network, a typical time series forecasting method.
	\item \textbf{DCRNN}\cite{DBLP:conf/iclr/LiYS018}: A Diffusion Convolutional Recurrent Neural Network that models spatial-temporal dependency by integrating graph convolution into gated recurrent units. 
	\item \textbf{STGCN}\cite{DBLP:conf/ijcai/YuYZ18}: Spatio-Temporal Graph Convolutional Networks that capture temporal dependency and spatial dependency by using 2D convolutional networks and graph convolutional network, respectively.
	\item \textbf{STG2Seq}\cite{DBLP:conf/ijcai/BaiYK0S19}: A Spatial-Temporal Graph to Sequence Model that uses multiple gated graph convolution module with two attention mechanisms to capture spatial-temporal dependency.
	\item \textbf{Graph WaveNet}\cite{DBLP:conf/ijcai/WuPLJZ19}: The Graph WaveNet combines graph convolution with dilated causal convolution to capture spatial-temporal dependency.
\end{itemize}
To enable a fair comparison, we use the same loss function in all models, defined as follows:
$$
Loss(\theta) = ||\mathbf{D}_{t+1} - \mathbf{\hat{D}}_{t+1}||_2^2,
$$
where $\mathbf{D}_{t+1},\mathbf{\hat{D}}_{t+1}$ denote the real request number and the predicted value, respectively. 
We normalize the context features data to the unit interval using Max-Min normalization and assign them to 10 clusters using the TICC algorithm. The request number is preprocessed using Mean-Std normalization.

For the route planning problem, we compare \textsf{DROP} with four baselines: 
\begin{itemize}
	\item \textbf{RD} \cite{giscup2019}: Random Destination randomly chooses a node as the destination and then uses the shortest-travel-path between the current location and the destination as the search route.
	\item \textbf{SmartAgent} \cite{DBLP:conf/gis/KimPZ19}: SmartAgent uses non-negative matrix factorization (NMF) to model and predict the spatio-temporal distributions of requests, and then chooses destinations using a greedy heuristic.
	\item \textbf{TripBandAgent} \cite{DBLP:conf/gis/BoruttaSF19}: TripBandAgent optimizes the taxi
	search strategy using reinforcement learning (RL).
	\item \textbf{STP} \cite{DBLP:conf/gis/HuMTZ19}: Spatial-Temporal Partitioning divides the search space into regions and computes weights for planning routes.
\end{itemize}

\begin{table}
	\centering
	\caption{Parameters (Default value is highlighted)}
	\label{tb:parameters}
	\resizebox{\linewidth}{!}{
		\begin{tabular}{|m{1.8cm}<{\centering}|l|}
			\cline{1-2}
			\textbf{Parameters} & \textbf{Values} \\
			\cline{1-2}
			Agent cardinality &\tabincell{l}{\textbf{5000}, 5500, 6000, 6500, 7000 (New York)\\ \textbf{1000}, 1100, 1200, 1300, 1400, 1500 (Haikou)}\\ 
			\cline{1-2}
			MLT (min)&5, 6, 7, 8, 9, \textbf{10}\\
			\cline{1-2}
			$\delta$&0.05, \textbf{0.1}, 0.15, 0.2, 0.25\\
			\cline{1-2}
			$\gamma$&0.6, \textbf{0.7}, 0.8, 0.9\\
			\cline{1-2}
			$\alpha$&0.5, \textbf{0.6}, 0.7, 0.8\\
			\cline{1-2}
			$L$&\tabincell{l}{4, 5, 6, \textbf{7} (New York)\\ 1, \textbf{2}, 3, 4 (Haikou)}\\
			\cline{1-2}
			$n$&2, 3, 4, \textbf{5}, 6 \\
			\cline{1-2}
			$\lambda$&0.3, 0.4, \textbf{0.5}, 0.6 \\
			\cline{1-2}
			
			\hline
	\end{tabular}}
\end{table}

\textbf{Parameter Settings.}
For request forecasting, we set the duration of time slot $t$ to 5 and 10 minutes for New York and Haikou, respectively. The number of regions $N$ is 99 for New York and 30 for Haikou. The batch size of the random gradient descent is 32 and the dropout rate is 0.2. The number of filters in the first and second ST-Gated block are 32 and 64, respectively. We use the Adam optimizer with default learning rate 0.001 for training, and we multiply the learning rate with 0.7 every five epochs. Parameter $h$ is set to 10, while $k$ and $m$ are set to 3. Threshold $\epsilon$ is selected from $\{0.2,0.4,0.5,0.6,0.7,0.8,1.0 \}$.

For route planning, we summarize the parameter settings of \textsf{DROP} in Table \ref{tb:parameters}. When we vary a parameter, other parameters are set to their default values.

\textbf{Evaluation Metrics.} For the request forecasting problem, we use three well-adopted metrics: Mean Average Percentage Error (MAPE), Mean Absolute Error (MAE), and Rooted Mean Square Error (RMSE).
For the route planning problem, we use three evaluation metrics: 
\begin{enumerate}
	\item The average agent idle time (highest priority).
	\item The average request waiting time, i.e., the period of time from its introduction to its pick-up or expiration.
	\item The percentage of expired requests.
\end{enumerate}


\subsection{Request Forecasting Performance Evaluation} 

We first compare \textsf{ST-GCSL} with the baseline models. Then, we conduct an ablation study on \textsf{ST-GCSL}. Further, we analyze the effect of parameter $\epsilon$ and the cost of training. Finally, we compare our model with the baseline models at multi-step prediction. 
For the data of each month, we use the last $10$ days for validation and testing (i.e., 5 days for validation and 5 days for testing) and the remaining days for training. 

\begin{table}
	\centering 
	\caption{Comparison with Baseline Models}
	\label{tb:comparision_with_baselines}
	\resizebox{\linewidth}{!}{
		\begin{tabular}{c|c|c|c|c|c|c}
			\hline
			\hline
			\multirow{2}{*}{Method} &
			\multicolumn{3}{c|}{New York} &
			\multicolumn{3}{c}{Haikou} \\
			\cline{2-7} 
			& MAPE (\%) & MAE & RMSE & MAPE (\%) & MAE & RMSE\\
			\hline
			HA & 43.71 & 3.781 & 5.993  & 46.36 & 4.563 & 7.047 \\
			VAR & 38.35 & 3.143 & 4.775  & 37.37 & 3.505 & 5.363 \\
			LSTM & 22.04 & 1.938 & 3.662 & 37.12 & 2.783 & 4.713 \\
			DCRNN & 17.42 & 1.883 & 3.659 & 32.21 & 2.722 & 4.731 \\
			STGCN & 17.63 & 1.945 & 3.840 & 32.33 & 2.772 & 4.843 \\
			STG2Seq & 17.74 & 1.943 & 4.044 & 31.04 & 2.808 & 4.882 \\
			Graph WaveNet & 18.64 & 1.916 & 3.770 & 35.36 & 2.933 & 5.045 \\
			\hline
			\textbf{ST-GCSL} & \textbf{16.46} & \textbf{1.788} &\textbf{3.563} & \textbf{29.30}& \textbf{2.663} & \textbf{4.700} \\
			\hline
			\hline
	\end{tabular}}
\end{table}

\textbf{Comparison With Baseline Models.}
Table \ref{tb:comparision_with_baselines} shows the forecasting test errors for the different methods. We have the following observations: 
\begin{enumerate}
	\item The classical methods, including HA and VAR, exhibit poor performance. The reason is that they are incapable of modeling non-linear spatial-temporal dependencies.
	\item In general, the deep learning methods perform better. LSTM only takes temporal dependencies into consideration, while DCRNN, STGCN, STG2Seq, and Graph WaveNet use two modules to model temporal and spatial dependencies, respectively. So they perform better than LSTM. It is worth noting that Graph waveNet performs poorly on Haikou, which may be caused by data sparsity.
	\item \textsf{ST-GCSL} achieves the best performance across all metrics on both datasets. The method takes short-term and long-term spatial-temporal dependencies into account and can capture temporal, spatial, and spatial-temporal dependencies, while other methods disregard either spatial or spatial-temporal dependencies.
\end{enumerate}

\begin{table}
	\centering  
	\caption{Comparison with Variants of ST-GCSL} 
	\label{tb:comparision_with_variants} 
	\resizebox{\linewidth}{!}{
		\begin{tabular}{c|c|c|c|c|c|c}
			\hline
			\hline
			\multirow{2}{*}{Removed Component} &
			\multicolumn{3}{|c|}{New York} &
			\multicolumn{3}{|c}{Haikou} \\
			\cline{2-7}   
			& MAPE (\%) & MAE & RMSE & MAPE (\%) & MAE & RMSE\\
			\hline
			Context features & 16.63 & 1.809 & 3.662 & 29.95 & 2.704 & 4.809\\
			GLU & 17.96 & 1.809 & 3.553 & 31.595& 2.703 & 4.716 \\
			STCM  & 17.13 & 1.792 & \textbf{3.543}& 29.65 & 2.674 & 4.673\\
			MSTCM & 16.77 & 1.788 & 3.549 & 30.02 & 2.677 & \textbf{4.664} \\
			Semantic-Neighbor & 16.77 & 1.795 & 3.572 & 29.33 &2.663 & 4.700 \\
			Geographical-Neighbor & 16.51 & 1.789 & 3.572 & 29.36 & 2.663 & 4.699 \\
			\hline
			\textbf{ST-GCSL} & \textbf{16.46} & \textbf{1.788} & 3.563 & \textbf{29.30}& \textbf{2.663} & 4.700 \\
			\hline
			\hline
	\end{tabular}}
\end{table}

\textbf{Ablation Study.}
To evaluate the effect of the different components of our model, we compare different variants of \textsf{ST-GCSL}, including: 
\begin{enumerate}
	\item Removing the context features.
	\item Replacing the GLU with the ReLU activation function.
	\item Removing the outer-STCM from ST-Gated blocks.
	\item Removing the MSTCM from ST-Gated block.
	\item Removing semantic neighbors from the region correlation graph.
	\item Removing geographical neighbors from the region correlation graph.
\end{enumerate}
The results are shown in Table \ref{tb:comparision_with_variants}. We have four observations. 
\begin{enumerate}
\item Without the context features, the model has poor performance, indicating that the context features contain information useful for prediction.
\item The model with the GLU performs better than the model with the ReLU activation function. This is because the module with GLU has twice as many parameters as ReLU, enabling it capture more complex spatial-temporal dependencies. Also, the gate in GLU is better at controlling the output than is ReLU.
\item Removing the outer-STCM or the MSTCM from ST-Gated Blocks may disregard temporal information and spatial-temporal dependencies to some extent. Specifically, removing the outer-STCM may disregard long-term temporal dependencies while removing the MSTCM may miss short-term spatial-temporal dependencies. Although removing the STCM and the MSTCM yields slightly better performance in RMSE, it is normal to obtain a slight change of results due to data sparsity.
\item Removing one type of neighbors weakens the performance, indicating that it is valuable to consider two types of neighbors.
\end{enumerate}
Overall, the results indicate that our proposed network structure is capable of outperforming the competitors.

\textbf{Effect of Parameter $\epsilon$.}
According to the results in Table \ref{tb:comparision_with_variants}, it is beneficial to consider semantic neighbors. Since parameter $\epsilon$ controls the number of semantic neighbors considered, we study the effect of $\epsilon$. Table \ref{tb:mape_with_epsilon} shows the prediction error with respect to $\epsilon$. We can see that when $\epsilon$ is 0.7 and 0.5, \textsf{ST-GCSL} achieves the best performance on New York and Haikou. The reason is that it considers many regions with weak dependency when $\epsilon$ is smaller; and when $\epsilon$ is larger, it only considers the regions with large similarity, which yields results similar to those obtained when disregarding semantic neighbors. Therefore, $\epsilon$ is set to a medium-large value.

\begin{table}
	\centering  
	\caption{MAPE When Varying $\epsilon$} 
	\label{tb:mape_with_epsilon} 
	\resizebox{\linewidth}{!}{
		\begin{tabular}{c|c|c|c|c|c|c|c}
			\hline
			\hline
			\multirow{2}{*}{Dataset} &
			\multicolumn{7}{c}{MAPE (\%)} \\
			\cline{2-8}   
			
			& 0.2 & 0.4 & 0.5 & 0.6 & 0.7 & 0.8 & 1.0 \\
			\hline
			\hline
			New York & 16.641 & 16.610 & 16.641  & 16.648 & \textbf{16.461}& 16.646 & 16.711\\
			\hline
			Haikou & 29.383 & 29.583 & \textbf{29.301} & 29.725 & 30.021& 30.048& 30.206\\
			\hline
			\hline
	\end{tabular}}
\end{table}

\begin{table}
	\centering  
	\caption{Training Time Comparison} 
	\label{tb:comparision_for_training_time} 
	\resizebox{\linewidth}{!}{
		\begin{tabular}{c|c|c|c|c|c}
			\hline
			\hline
			\multirow{2}{*}{Dataset} &
			\multicolumn{5}{c}{Time Consumption (s)} \\
			\cline{2-6}   
			
			& DCRNN & STGCN & STG2Seq & Graph WaveNet & \textbf{ST-GCSL}\\
			\hline
			\hline
			New York & 1616.8 & 124.5 & 705.4  & 930.0 & 470.9\\
			\hline
			Haikou & 561.4 & 99.5 & 215.4 & 297.2 & 173.8\\
			\hline
			\hline
	\end{tabular}}
\end{table}

\begin{table*}
	\centering 
	\caption{Multi-step Prediction}
	\label{tb:multi-step_predicted}
	\resizebox{.8\linewidth}{!}{
		\begin{tabular}{c|c|c|c|c|c|c}
			\hline
			\hline
			\multirow{2}{*}{Method} &
			\multicolumn{3}{c|}{New York (step 2/3)} &
			\multicolumn{3}{c}{Haikou (step 2/3)} \\
			\cline{2-7}   
			& MAPE (\%) & MAE & RMSE & MAPE (\%) & MAE & RMSE\\
			\hline
			DCRNN & 17.83/18.44 & 1.923/1.964 & 3.763/3.847 & 33.46/35.58 & 2.812/2.966 & 4.906/5.210 \\
			STGCN & 18.02/18.51 & 2.012/2.075 & 4.032/4.197 & 32.38/32.97 & 2.860/3.005 & 4.986/5.244 \\
			STG2Seq & 17.75/17.83 & 2.124/2.315 & 4.665/5.277 & 32.21/33.92 & 2.991/3.221 & 5.222/5.631 \\
			Graph WaveNet & 20.19/21.05 & 2.022/2.111 & 4.130/4.336 & 40.44/42.61 & 3.122/3.314 & 5.322/5.637 \\
			\textbf{ST-GCSL} & \textbf{16.71}/\textbf{16.95} & \textbf{1.835}/\textbf{1.876} &\textbf{3.693}/\textbf{3.801} & \textbf{29.67}/\textbf{30.67}& \textbf{2.764}/\textbf{2.922} & \textbf{4.895}/\textbf{5.194} \\
			\hline
			\hline
	\end{tabular}}
\end{table*}

\textbf{Training Time Analysis.}
We compare \textsf{ST-GCSL} with the baseline models with respect to training time. For fairness, all models are trained with the best parameters in 100 epochs. The results in Table \ref{tb:comparision_for_training_time} indicate the following. 
\begin{enumerate}
\item STGCN is the fastest at training on both datasets, due to the convolution along time axis and minimal GC operations.
\item DCRNN is the slowest due to its integration of GC into the gated recurrent unit, which means that the GC operation is executed at each time step.
\item \textsf{ST-GCSL} achieves the second-best training time performance. Although the parallel operation of MSTCM in \textsf{ST-GCSL} is similar with that of STG2Seq, the design of outer-STCM decreases the number of GC operations, which makes \textsf{ST-GCSL} more efficient. Unlike STG2Seq and \textsf{ST-GCSL}, Graph WaveNet needs to learn a self-adaptive adjacency matrix to determine unseen graph structures, which takes extra time.
\end{enumerate}
The experimental findings thus indicate that \textsf{ST-GCSL} not only achieves the best performance, but also can be trained efficiently.

\textbf{Multi-step Prediction Comparison.}
\textsf{ST-GCSL} is able to conduct multi-step prediction, as can DRCNN, STGCN, STG2Seq, and Graph WaveNet. So we compare \textsf{ST-GCSL} with these methods at request forecasting in 2 and 3 time steps. Table \ref{tb:multi-step_predicted} presents the results for all metrics on both datasets. As we can see, ST-GCSL exhibits the best performance.

\begin{figure*}
	\centering
	\subfigure{\includegraphics[width=.5\linewidth]{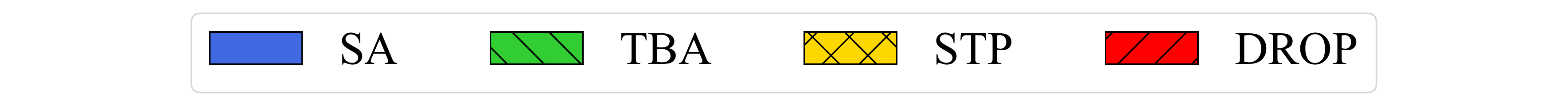}}
	\\
	\setcounter{subfigure}{0}
	\subfigure[Idle Time]{\includegraphics[width=.3\linewidth]{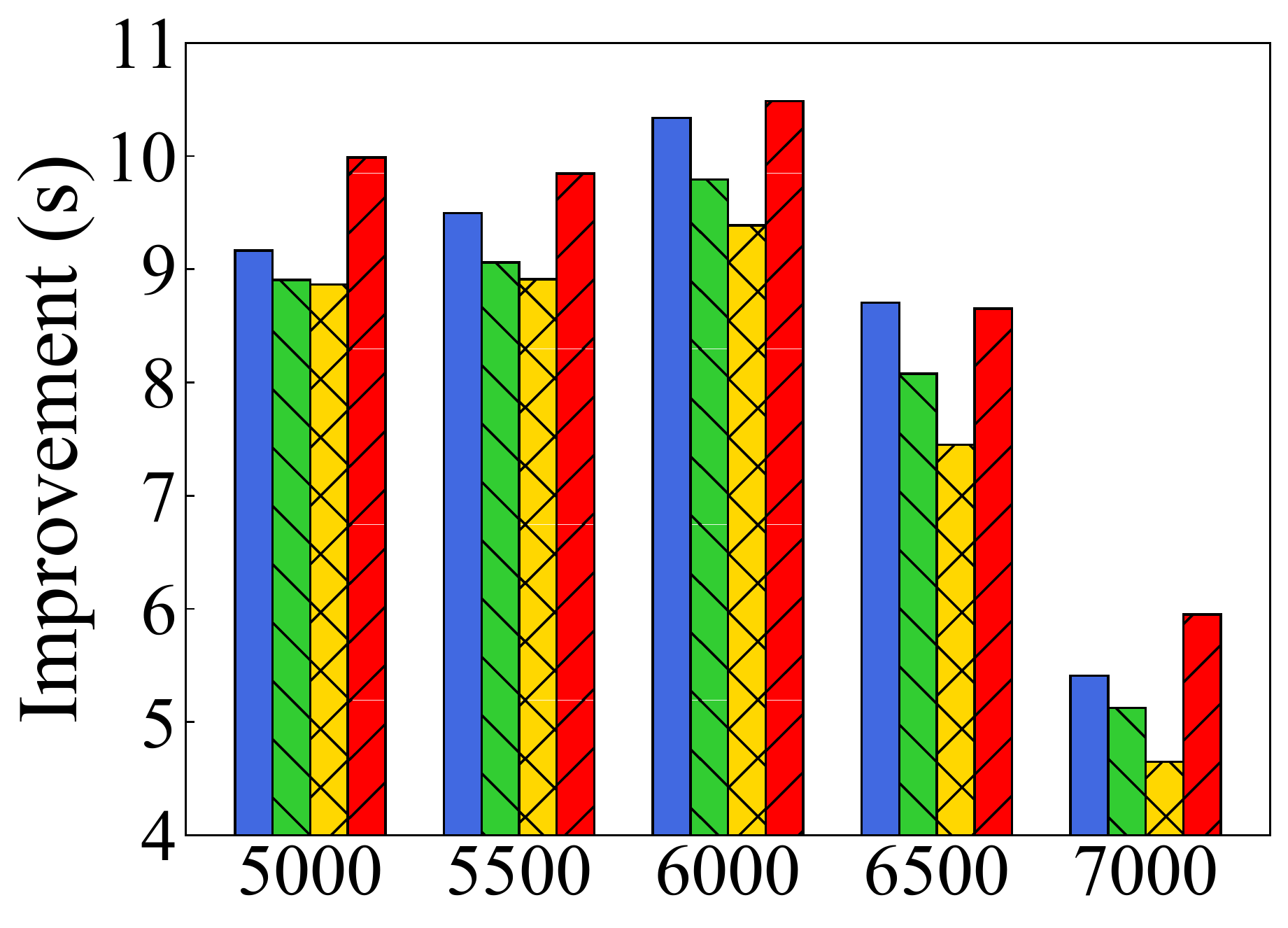}\label{expe:search_bar}}
	\subfigure[Waiting Time]{\includegraphics[width=.3\linewidth]{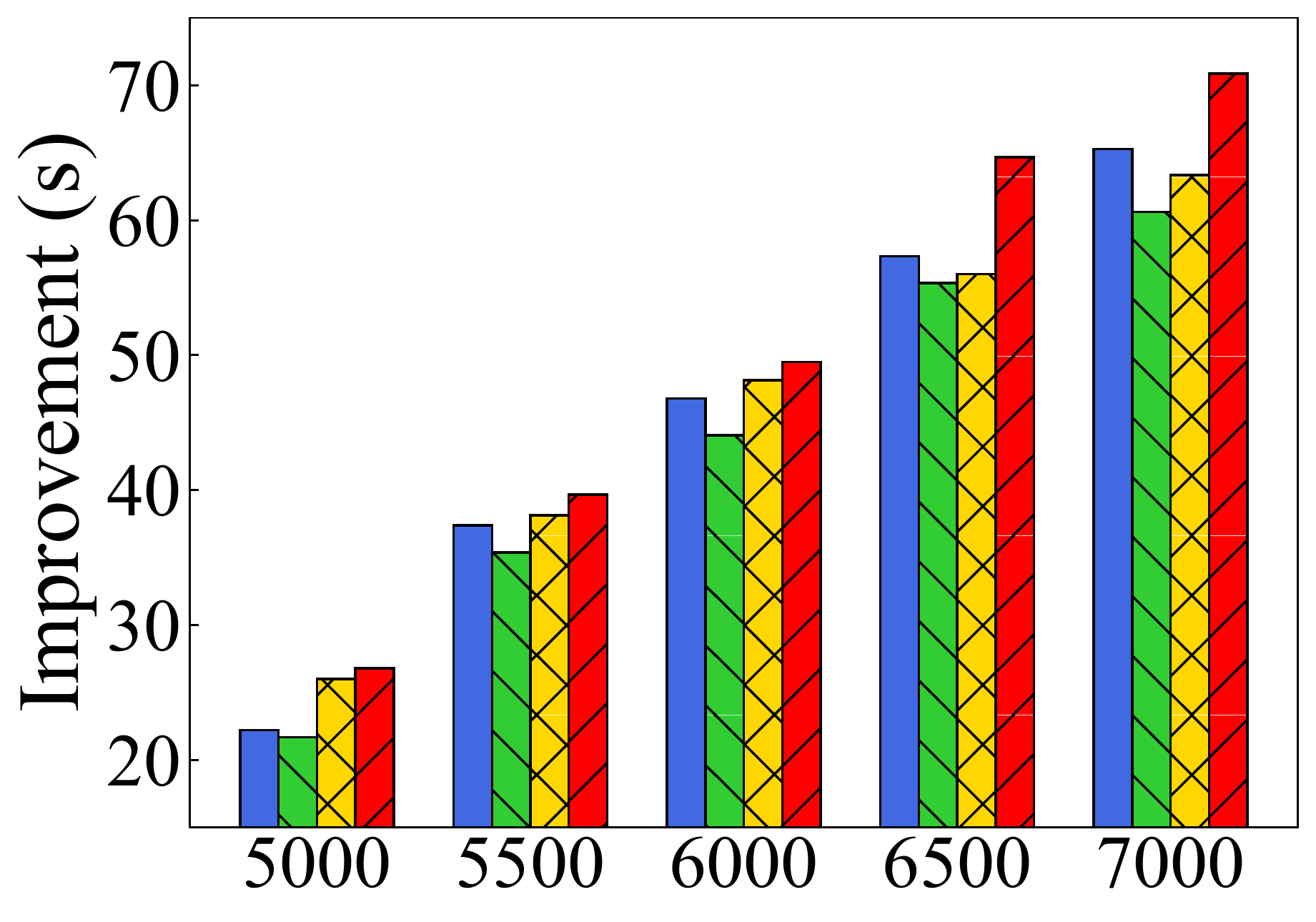}\label{fig:mape_haikou}}
	\subfigure[Expiration Percentage]{\includegraphics[width=.3\linewidth]{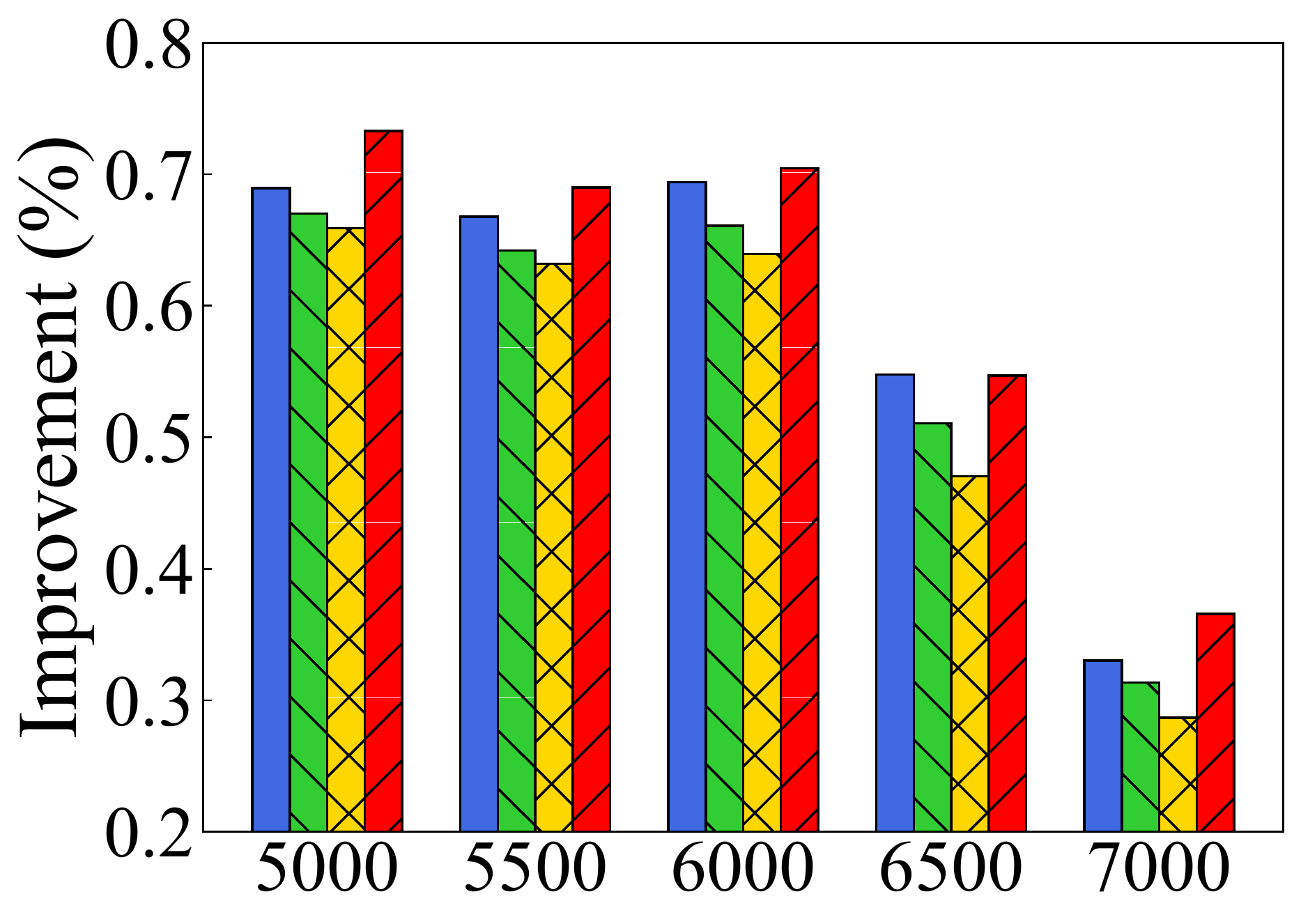}\label{fig:rmse_haikou}}
	\caption{Comparison with Baselines on New York Varying the Agent Cardinality}\label{expe:NYC_AC}
\end{figure*}

\begin{figure*}
	\centering
	\subfigure[Idle Time]{\includegraphics[width=.3\linewidth]{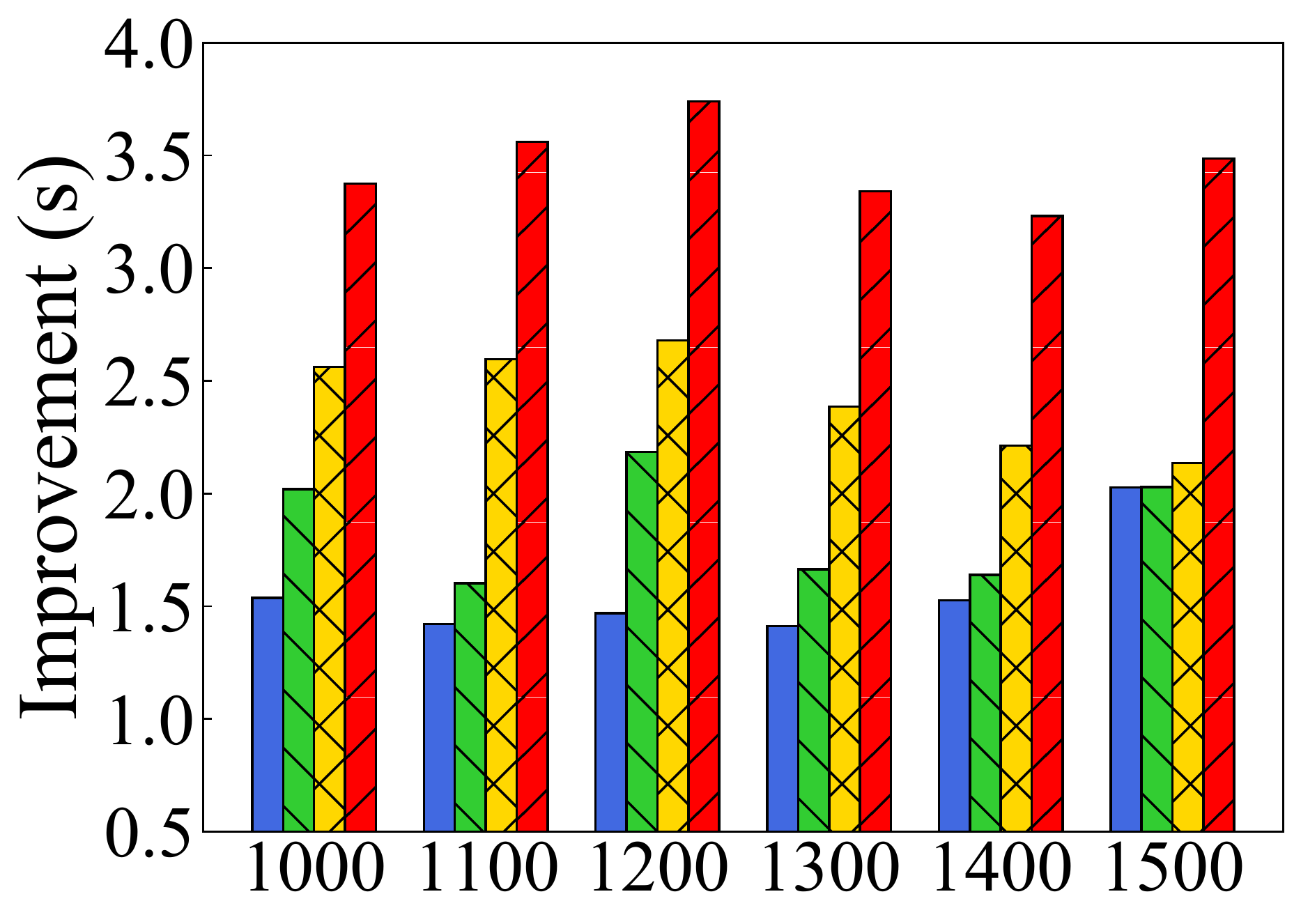}\label{expe:search_haikou}}
	\subfigure[Waiting Time]{\includegraphics[width=.3\linewidth]{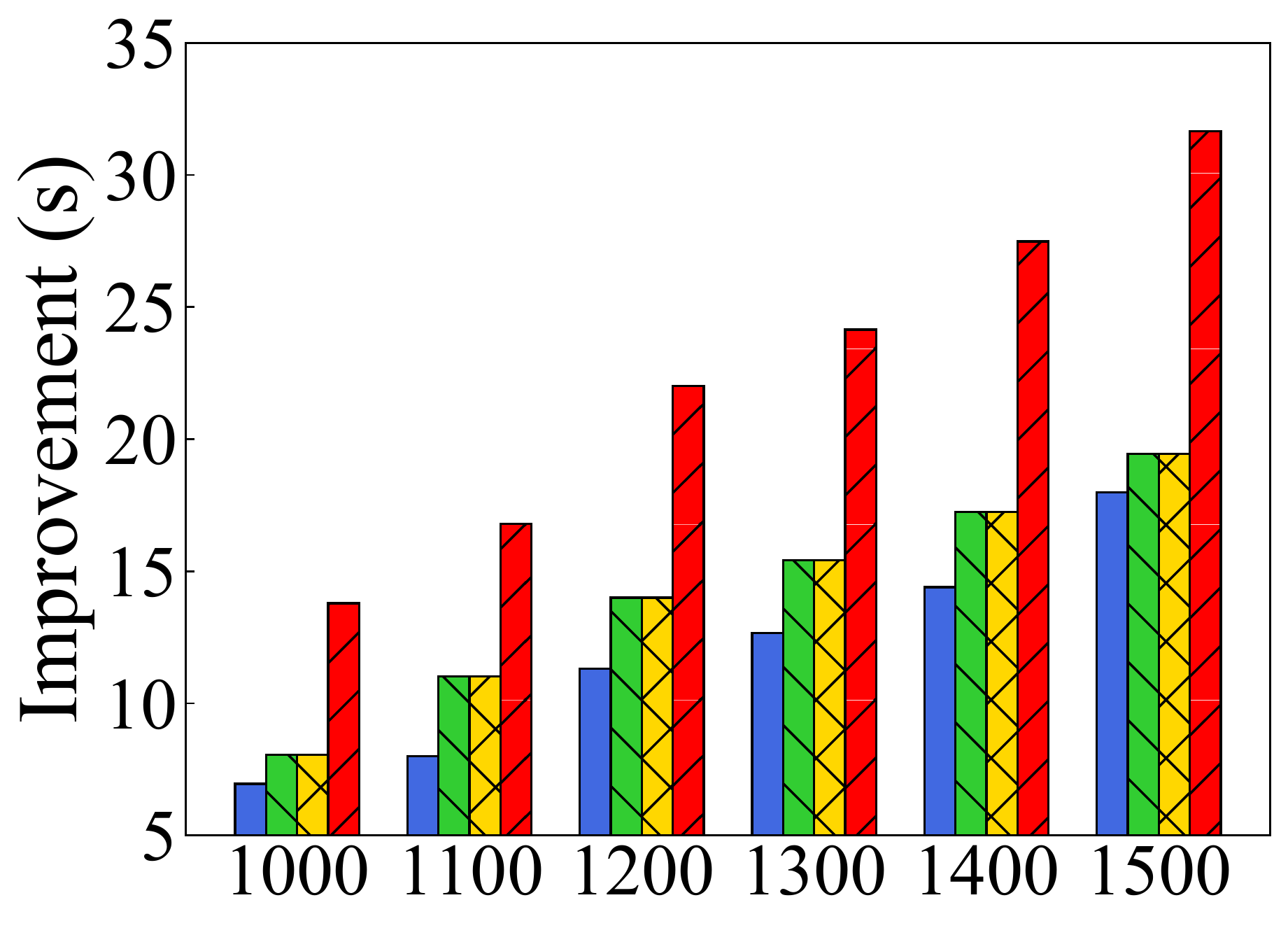}\label{expe:waiting_haikou}}
	\subfigure[Expiration Percentage]{\includegraphics[width=.3\linewidth]{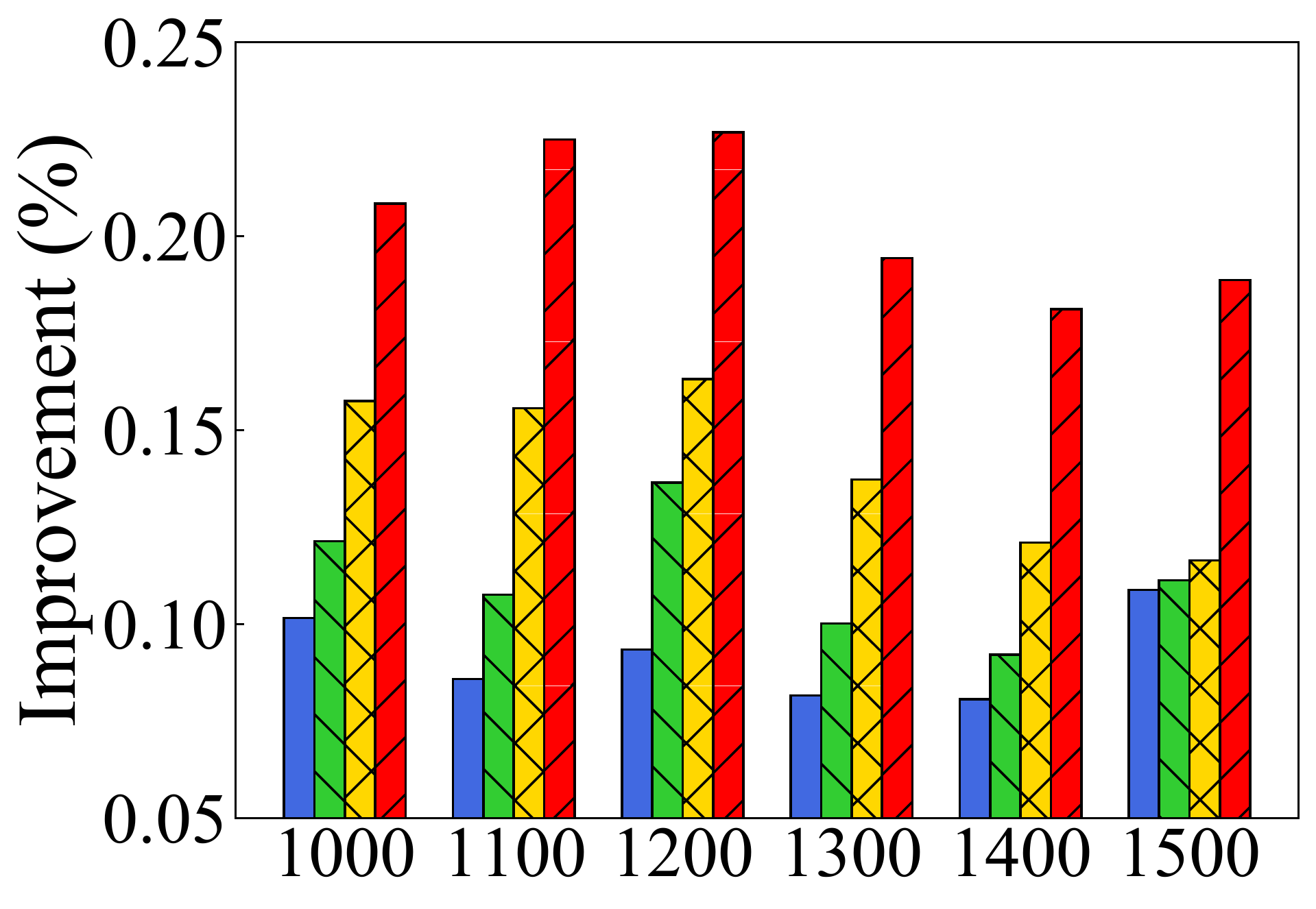}\label{expe:expiration_haikou}}
	\caption{Comparison with Baselines on Haikou Varying the Agent Cardinality}\label{expe:Haikou_AC}
\end{figure*}

\begin{figure*}
	\centering
	\subfigure[Idle Time]{\includegraphics[width=.3\linewidth]{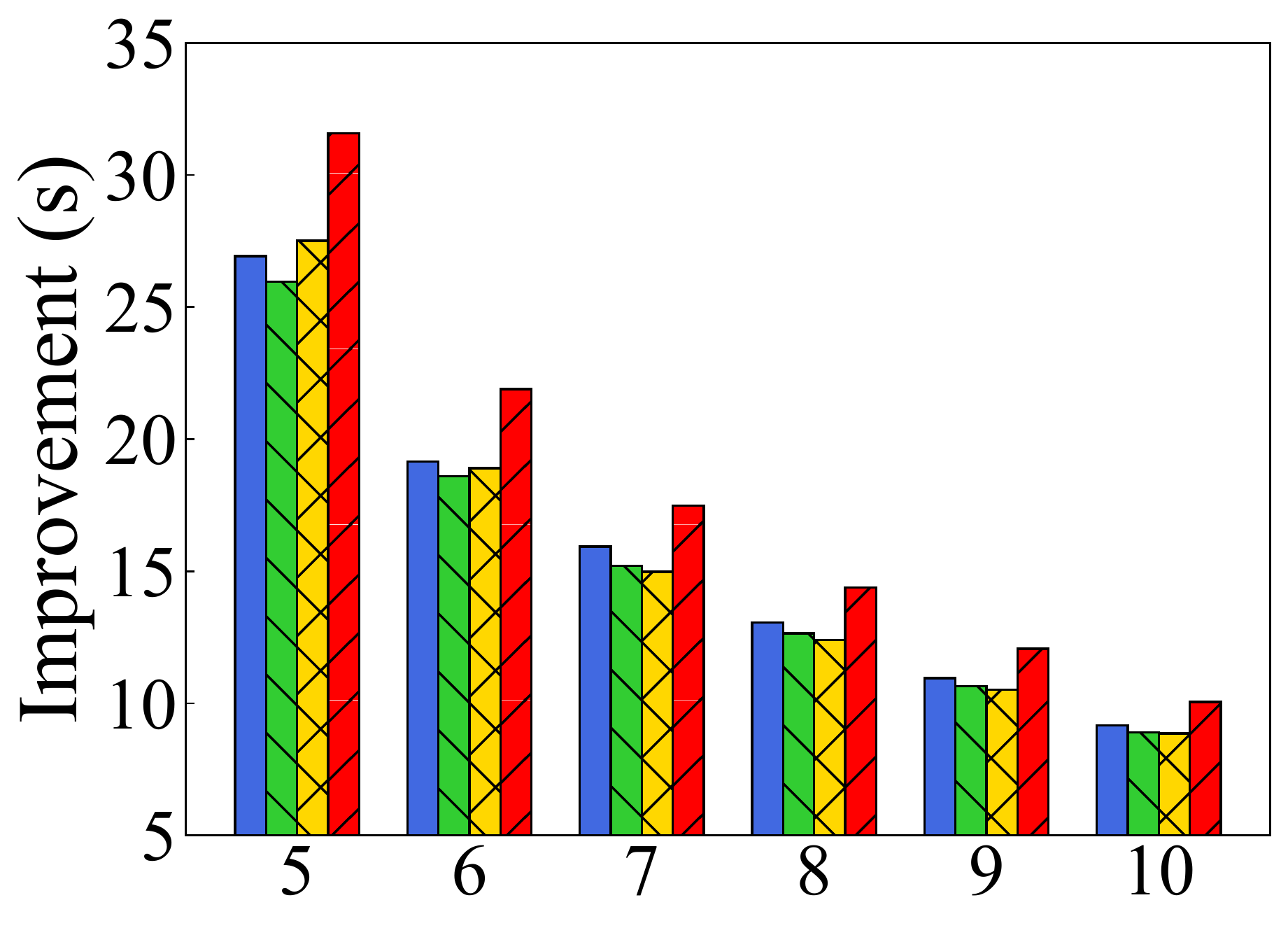}\label{expe:search_mlt}}
	\subfigure[Waiting Time]{\includegraphics[width=.3\linewidth]{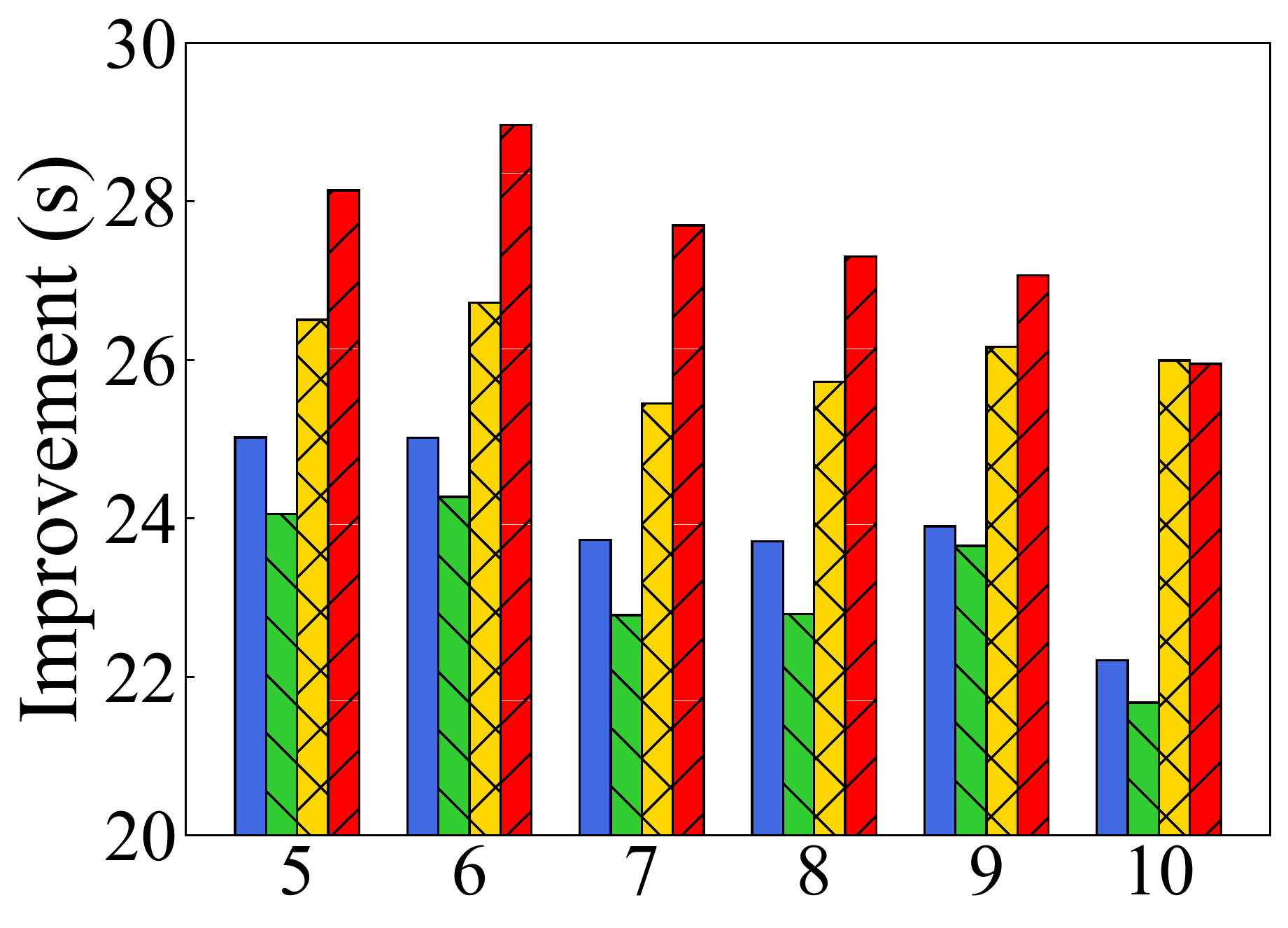}\label{expe:waiting_mlt}}
	\subfigure[Expiration Percentage]{\includegraphics[width=.3\linewidth]{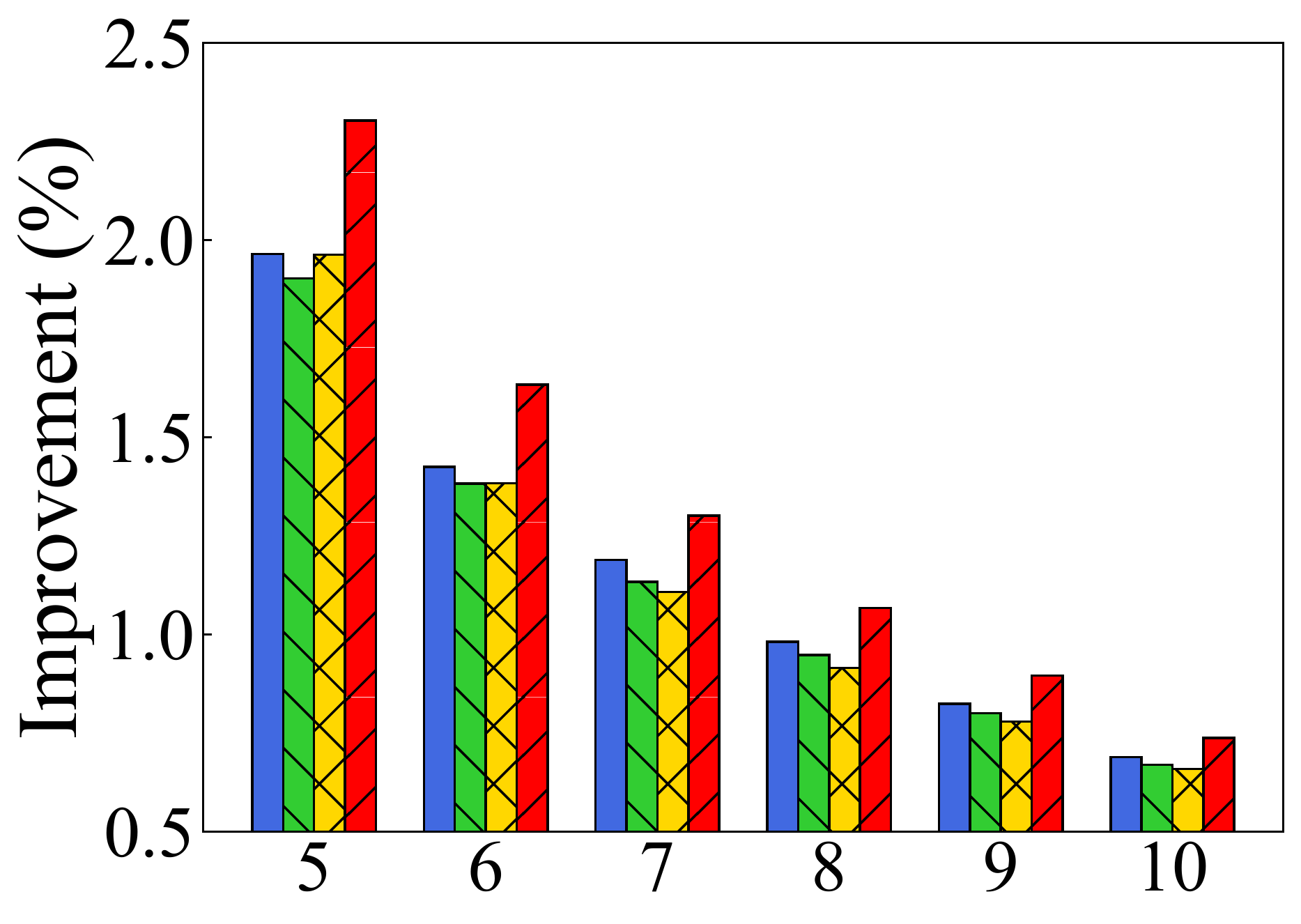}\label{fig:expiration_mlt}}
	\caption{Comparison with Baselines on New York Varying the MLT}\label{expe:NYC_MLT}
\end{figure*}

\begin{figure*}
	\centering
	\subfigure[Idle Time]{\includegraphics[width=.3\linewidth]{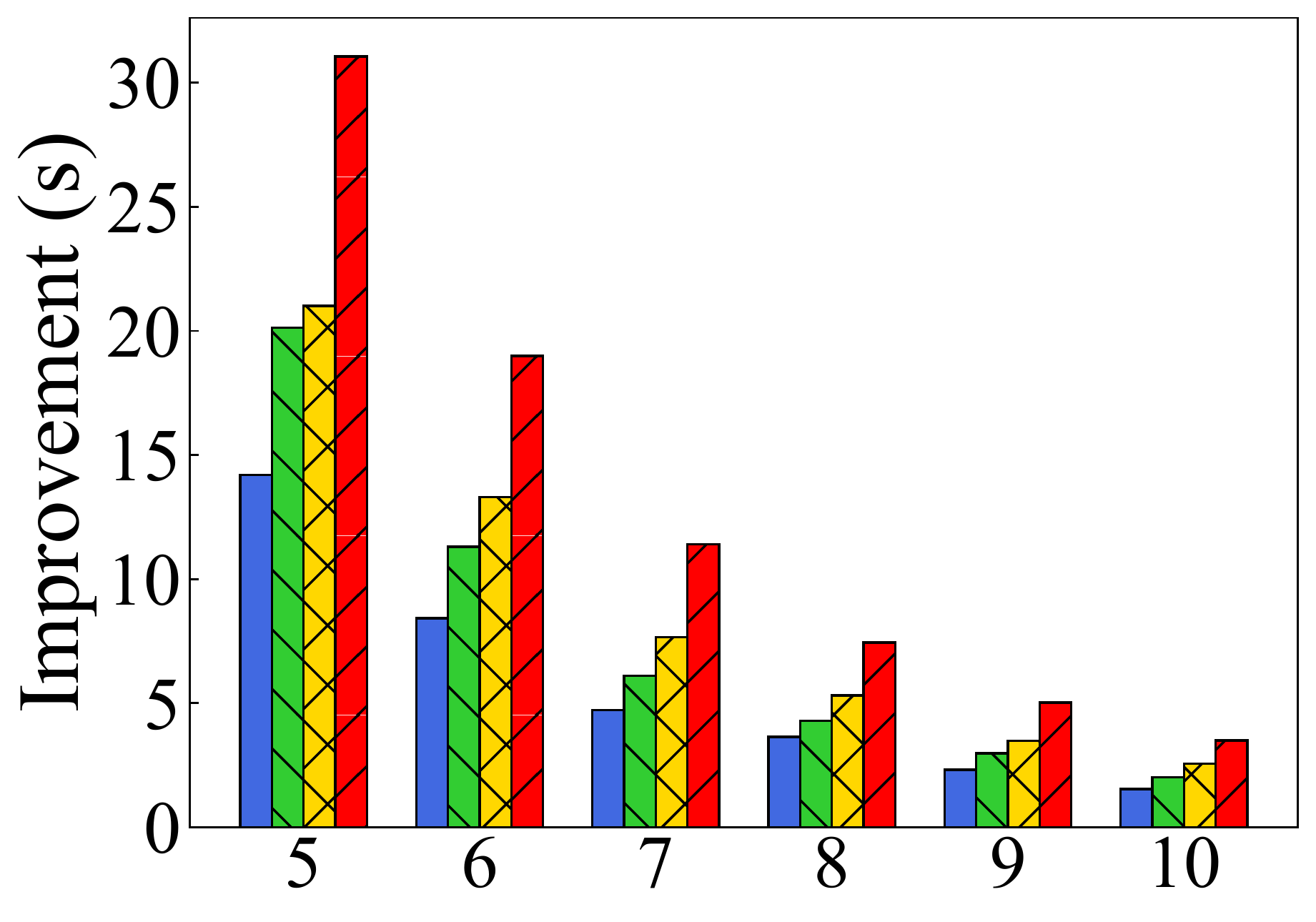}\label{expe:search_haikou_mlt}}
	\subfigure[Waiting Time]{\includegraphics[width=.3\linewidth]{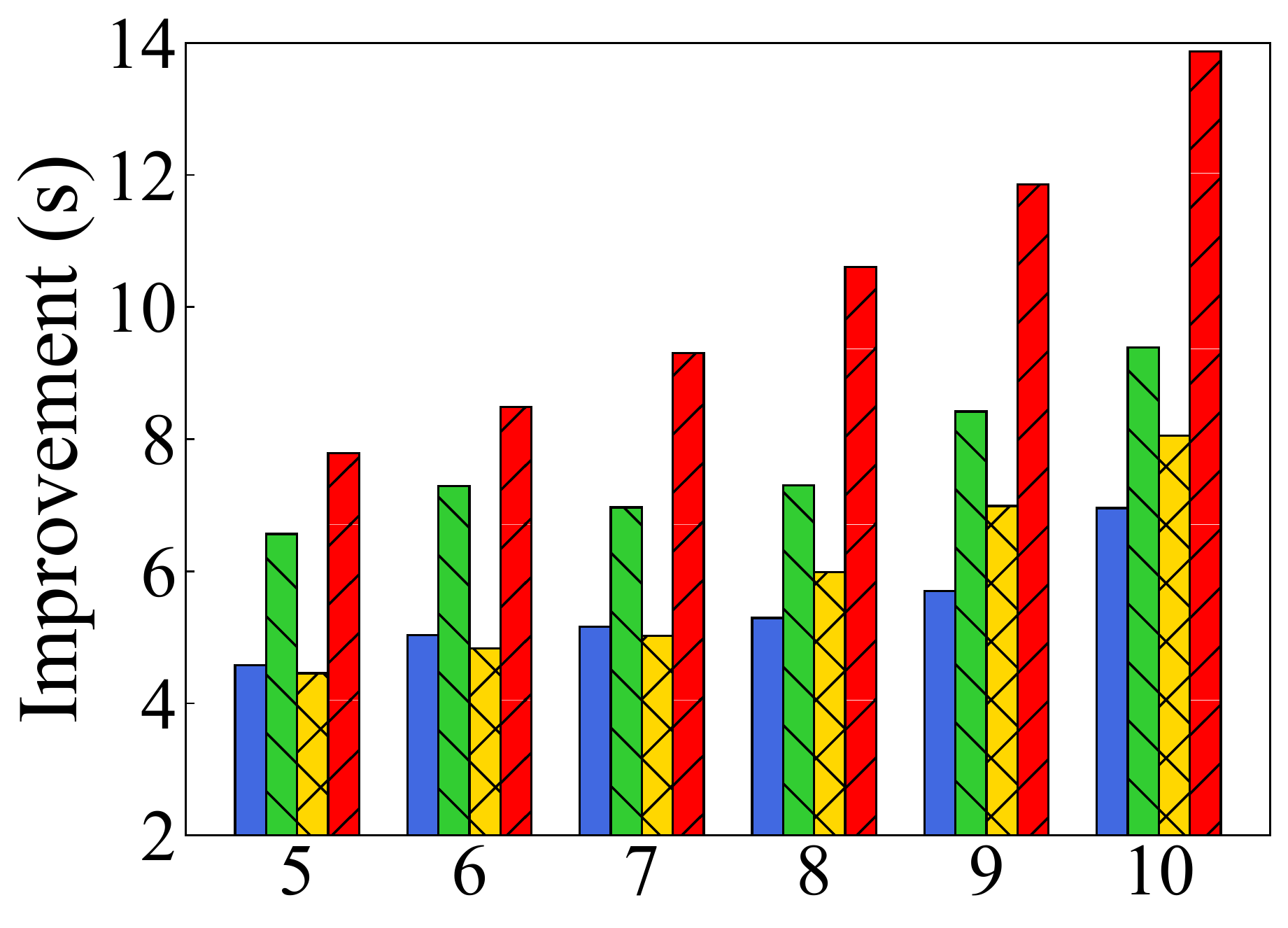}\label{expe:waiting_haikou_mlt}}
	\subfigure[Expiration Percentage]{\includegraphics[width=.3\linewidth]{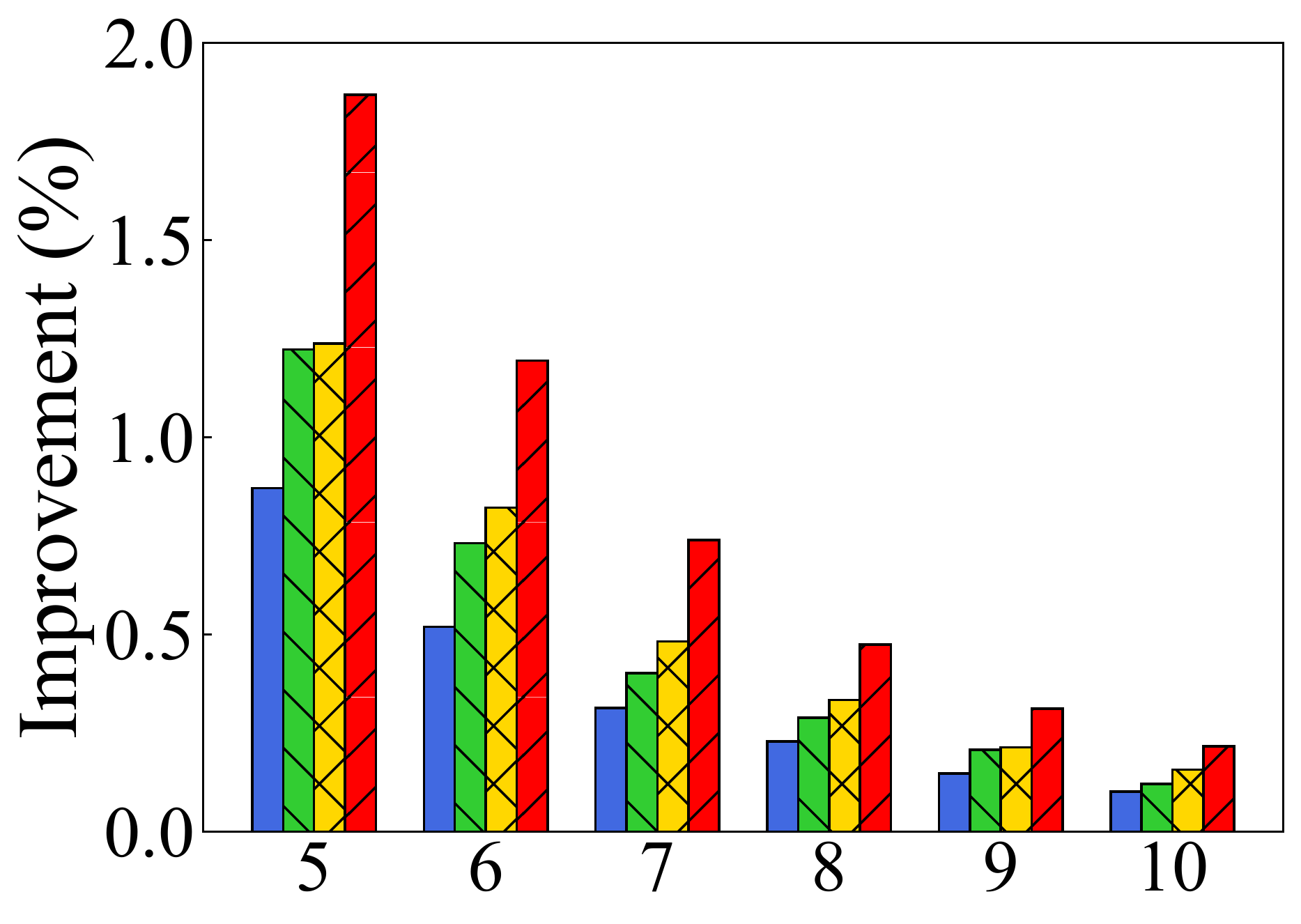}\label{fig:expiration_haikou_mlt}}
	\caption{Comparison with Baselines on Haikou Varying the MLT}\label{expe:Haikou_MLT}
\end{figure*}

\subsection{Route Planning Performance Evaluation} 

We randomly select two days data from each month to form the testing dataset, so we have 12 days data for evaluation. To better show the performance improvement, we report the average improvement compared to \textsf{RD} in Figs. \ref{expe:NYC_AC} to \ref{expe:Haikou_MLT} on previously mentioned three metrics. The experimental results of different parameter settings are shown in Table \ref{tb:parameter_expe}.

\textbf{Varying the Agent Cardinality.} Figs. \ref{expe:NYC_AC} and \ref{expe:Haikou_AC} show the performance of all methods when varying the agent cardinality. We make three observations. 
\begin{enumerate}
\item The idle time improvement of \textsf{DROP} is the highest. For instance, when the agent cardinality is 5000 on New York, the average idle time of RD is 368.8 seconds, and the improvement of the idle time for SA, TBA, and STP are 9.2, 8.9, and 8.8 seconds, while the improvement achieved by \textsf{DROP} is 10 seconds. The reason is that the predicted results of \textsf{ST-GCSL} are close to the real request distribution, and \textsf{DROP} guides agents to popular regions.
\item \textsf{DROP} outperforms the baseline methods in terms of request waiting time. With the increase of agent cardinality, the improvement of \textsf{DROP} increases compared to those of the other methods. This is because \textsf{DROP} can guide more agents to locations near requests.
\item \textsf{DROP} has the highest expiration percentage improvement compared to the baseline methods in all cases, since it distributes agents according to the distribution of requests.
\end{enumerate}

\textbf{Varying the MLT.} Figs. \ref{expe:NYC_MLT} and \ref{expe:Haikou_MLT} show the performance of all methods when varying the MLT. We make the following observations. 
\begin{enumerate}
\item \textsf{DROP} outperforms the other methods with respect to idle time, especially when MLT is small. For example, when MLT is 5 minutes on New York, the average idle time of RD is 400 seconds, and the improvement of \textsf{DROP} is 31.6 seconds, while those of SA, TBA, and STP are 26.9, 26, and 27.5 seconds, respectively. This is because when MLT is small, the requests can only be assigned to nearby agents, and the routing strategy of \textsf{DROP} can direct agents to locations closer to requests than can the other methods.
\item \textsf{DROP} has the best performances in terms of waiting time and expiration percentage on the two datasets.
\end{enumerate}

\textbf{Parameter Study.} We compare the performance of \textsf{DROP} under the different parameter settings shown in Table \ref{tb:parameter_expe}. We observe that the performance changes only little across these settings. Note that as the two datasets are of different sizes, there are differences in the parameter settings. For instance, the size of the neighbor order $L$ on New York varies from 4 to 7, while $L$ varies from 1 to 4 on Haikou. The optimal value of $L$ is 7 on New York, while it is 3 on Haikou.

\begin{table}[htbp]
	\centering
	\caption{Performance of Different Parameter Settings}
	\vspace*{0.1in}
	\label{tb:parameter_expe}
	\resizebox{\linewidth}{!}{
		\begin{tabular}{|c|c|m{1cm}<{\centering}|m{1cm}<{\centering}|m{1cm}<{\centering}|m{1cm}<{\centering}|m{1cm}<{\centering}|m{1cm}<{\centering}|}
			\cline{1-8}
			
			\multicolumn{2}{|c|}{}  & \multicolumn{3}{|c|}{New York} & \multicolumn{3}{|c|}{Haikou}\\
			\cline{1-8}
			
			\multicolumn{2}{|c|}{Parameters}  & IT (s) & WT (s) & Exp (\%) & IT (s) & WT (s) & Exp (\%) \\
			\hline
			
			\multirow{5}*{\textbf{$\delta$}}&0.05&358.962&\textbf{249.699}&\textbf{7.944}&\textbf{436.259}&\textbf{292.485}&\textbf{8.297}\\	
			\multicolumn{1}{|c|}{}&0.1&\textbf{358.826}&250.464&7.946&436.293&292.831&8.303\\
			\multicolumn{1}{|c|}{}&0.15&358.936&252.038&7.953&436.636&294.172&8.323\\
			\multicolumn{1}{|c|}{}&0.2&359.087&255.807&7.963&436.957&	295.317&	8.344\\
			\multicolumn{1}{|c|}{}&0.25&359.430&	257.495&	7.984&	437.451&	297.965&	8.375\\
			\cline{1-8}

			\multirow{4}*{\textbf{$\gamma$}}&0.6&358.959&	\textbf{250.094}&	7.946&	\textbf{436.121}&	\textbf{292.133}&	\textbf{8.288}	\\	
			\multicolumn{1}{|c|}{}&0.7&358.826&	250.464&	7.946&	436.259&	292.485&	8.297			\\
			\multicolumn{1}{|c|}{}&0.8&358.897& 252.157&	7.951&	436.352&	292.185&	8.313\\
			\multicolumn{1}{|c|}{}&0.9&\textbf{358.776}&	250.348&	\textbf{7.944}&	436.570&	292.229&	8.321\\
			\cline{1-8}

			\multirow{4}*{\textbf{$\alpha$}}&0.5&358.978&	251.864&	7.957&	436.197&	\textbf{291.707}	&8.293\\	
			\multicolumn{1}{|c|}{}&0.6&\textbf{358.776}&	\textbf{250.348}&	7.944&	436.121&	292.133&	8.288\\
			\multicolumn{1}{|c|}{}&0.7&358.935&	252.499&	7.947&	\textbf{436.029}&	292.286&	\textbf{8.281}\\
			\multicolumn{1}{|c|}{}&0.8&358.843&	250.535	&\textbf{7.942}&	436.218&	292.608&	8.300\\
			\cline{1-8}
			
			\multirow{4}*{L}&4/1&359.937&	251.227&8.027&	436.938&	\textbf{291.620}	&8.341\\	
			\multicolumn{1}{|c|}{}&5/2&359.288&	250.707&	7.984&	436.029 &292.286	&8.281\\
			\multicolumn{1}{|c|}{}&6/3&359.120&	252.104&   7.957&	\textbf{435.681}&	293.111	&\textbf{8.262}\\
			\multicolumn{1}{|c|}{}&7/4&\textbf{358.776}&	\textbf{250.348}&	\textbf{7.944}&	435.793&	293.447	&8.270\\
			\cline{1-8}
			
			\multirow{5}*{n}&2&358.893&	251.031&	7.952&	435.716&	293.142	&\textbf{8.260}\\	
			\multicolumn{1}{|c|}{}&3&358.895&	250.733&	7.948&	435.745&	292.949	&8.265\\
			\multicolumn{1}{|c|}{}&4&358.870&	252.052&	7.951&	435.837&	293.003	&8.288\\
			\multicolumn{1}{|c|}{}&5&\textbf{358.776}&	\textbf{250.348}&	\textbf{7.944}&	\textbf{435.681}&	293.111&	8.262
			\\
			\multicolumn{1}{|c|}{}&6&359.031&	252.329&	7.957&	435.689&	\textbf{292.948}	&8.271\\
			\cline{1-8}

			\multirow{4}*{\textbf{$\lambda$}}&0.3&358.911&	251.401&	7.947&	436.096	&293.893	&8.298\\	
			\multicolumn{1}{|c|}{}&0.4&358.989&	251.029&	7.947&	436.032&	293.408	&8.280\\
			\multicolumn{1}{|c|}{}&0.5&\textbf{358.776}&	\textbf{250.348}&	\textbf{7.944}&	435.681	&293.111&	8.262\\
			\multicolumn{1}{|c|}{}&0.6&358.913&	250.636&	7.945&	\textbf{435.614}&	\textbf{292.061}	&\textbf{8.253}
			\\
			\cline{1-8}
			\hline
	\end{tabular}}
\end{table}

\section{Related Work}\label{sec:related}

\subsection{Taxi Demand Prediction}
Traffic demand prediction is a critical aspect when aiming to achieve an efficient transportation system. Many studies employ convolutional neural networks (CNNs). For example, ConvLSTM \cite{DBLP:conf/nips/ShiCWYWW15} combines CNNs and recurrent neural networks (RNNs) to model spatial and temporal dependencies, respectively, which is an extension of fully-connected LSTMs \cite{DBLP:journals/neco/HochreiterS97}. ST-ResNet \cite{DBLP:journals/ai/Zhang0QLYL18} models the temporal closeness, period, and trend properties of crowd traffic based on existing studies \cite{DBLP:conf/gis/ZhangZQLY16, DBLP:conf/aaai/ZhangZQ17}. DMVST-Net \cite{DBLP:conf/aaai/Yao0KTJLGYL18} employs graph embedding as an external feature to improve forecast accuracy based on localized spatial and temporal views. 

However, these CNN-based methods only model the Euclidean relations among grid regions and ignore non-Euclidean relations. In contrast, graph convolutional networks (GCNs) can extract local features from non-Euclidean structures, resulting in improved performance. For instance, One study \cite{DBLP:journals/corr/abs-1811-05157} predicts the travel cost, while GCWC \cite{DBLP:conf/icde/ChengCY19} fills in missing stochastic weights of speed via GCN, and their techniques can be applied to demand prediction. DCRNN \cite{DBLP:conf/iclr/LiYS018} models the spatial-temporal dependency by integrating graph convolution into the gated recurrent units. The architecture is like that of ConvLSTM, but replaces the Conv2d with GC. STGCN \cite{DBLP:conf/ijcai/YuYZ18} captures temporal dependency and spatial dependency by using 2D convolutional networks and graph convolutional networks, respectively. STG2Seq \cite{DBLP:conf/ijcai/BaiYK0S19} uses multiple gated graph convolution modules with two attention mechanisms to capture spatial-temporal dependency, while Graph WaveNet \cite{DBLP:conf/ijcai/WuPLJZ19} learns a self-adaptive adjacency matrix and combines graph convolution with dilated casual convolution to capture spatial-temporal dependencies. STMGCN \cite{DBLP:conf/aaai/ZhaoXSTZZ19} employs a multi-graph to extract spatial dependency and uses RNNs to capture temporal dependency. 

The GCN-based methods are more flexible and progressive than the CNN-based methods, but there are also differences between them. STGCN, Graph WaveNet, and STMGCN use separate components to capture temporal and spatial dependency, while the GC operation in DCRNN and STG2Seq is executed by short-term time steps, meaning that they capture spatial and temporal dependency simultaneously, which also leads to much more computation. Further, based on the three temporal and spatial dependencies in the traffic forecasting problem we study, the above methods all miss spatial-temporal dependency to some extent. Unlike those methods, our method captures all three types of dependency and can be trained efficiently.

\subsection{Taxi Dispatching}
Existing studies on taxi dispatching generally concern either order matching or route recommendation. Order matching aims to match idle taxis with appropriate passengers, often to maximize global revenue. One study \cite{DBLP:conf/kdd/XuLGZLNLBY18} considers both instant passenger satisfaction and expected future gain in a unified decision-making framework, and then optimizes long-term platform efficiency through reinforcement learning. Another study \cite{DBLP:conf/kdd/TangQZWXMZY19} models the ride dispatching problem as a Semi Markov Decision Process and proposes Cerebellar Value Networks (CVNet) with a novel distributed state representation layer to learn the best dispatch policy. Moreover, some studies \cite{DBLP:journals/jors/BaiLAK14, DBLP:conf/icde/ChengCY19} introduce the idea of game theory into the matching problem and model the matching of taxis and passengers as a process to reach a stable Nash equilibrium. 

Considering route recommendation, existing proposals can be divided into combination optimization methods and deep reinforcement learning (DRL) methods. One study \cite{DBLP:conf/www/XuMLJLYLWQ20} models the driver repositioning task as a classical Minimum Cost Flow (MCF) problem and then solves it by combination optimization. MCF-FM \cite{DBLP:conf/gis/MingHDZ20} develops a continuous order dispatch strategy for effective fleet management. Next, DRL is used widely to deal with taxi dispatching and repositioning problems. For instance, MOVI \cite{DBLP:conf/infocom/OdaJ18} uses CNNs to extract supply and demand distribution features and adopts a distributed DQN policy to learn the best dispatch action for each vehicle. One study \cite{DBLP:conf/kdd/LinZXZ18} considers geographic context and collaborative context to remove invalid actions, proposing a contextual actor-critic multi-agent reinforcement learning algorithm to adapt to dynamic traffic. However, these methods disregard the real-time nature of supply and demand, which hurts their performance. In contrast, our framework guides idle taxis based on real-time supply and demand distributions. 


\section{Conclusion} \label{sec:conclusion}
We study the problem of spatial-temporal demand forecasting and competitive supply (SOUP). We propose the \textsf{ST-GCSL} model for request prediction and develop the \textsf{DROP} algorithm to guide idle agents. We report on experimental studies that show that \textsf{ST-GCSL} is capable of outperforming the baseline models considered. Specifically, \textsf{DROP} outperforms all the competitors and reduces the idle time, waiting time, and expiration percentage by 10s, 26.8s, and 0.73\% compared with the RD method on the New York dataset.

\bibliographystyle{abbrv}
\bibliography{myRef}

\begin{IEEEbiography}[{\includegraphics[width=1in,height=1.25in,clip,keepaspectratio]{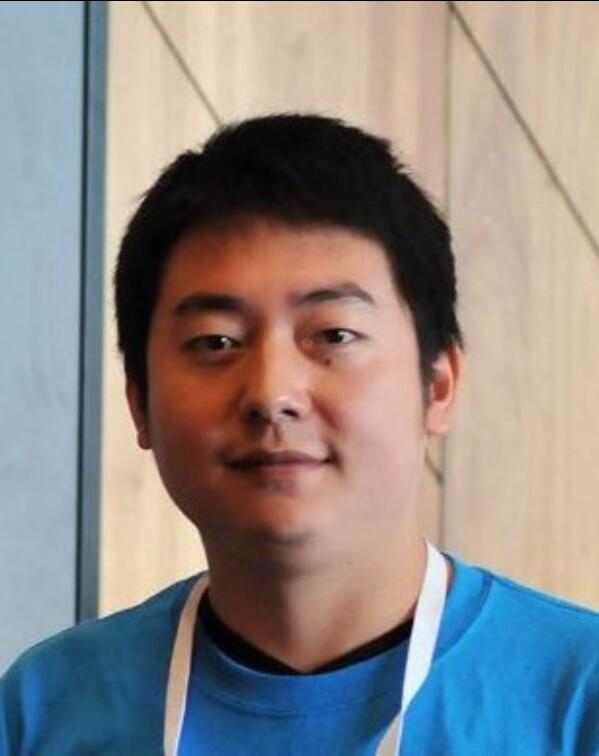}}]{Bolong Zheng} is an associate professor in Huazhong University of Science and Technology (HUST). He received his PhD from University of Queensland in 2017. 
He received bachelor and master degrees in computer science from HUST in 2011 and 2013, respectively. 
His research interests include spatial-temporal data management.
He has published over 40 papers in prestigious journals and conferences in data management field such as SIGMOD, ICDE, VLDB, ACM Transactions and IEEE Transactions. 
\end{IEEEbiography}

\vspace{-42pt}

\begin{IEEEbiography}[{\includegraphics[width=1in,height=1.25in,clip,keepaspectratio]{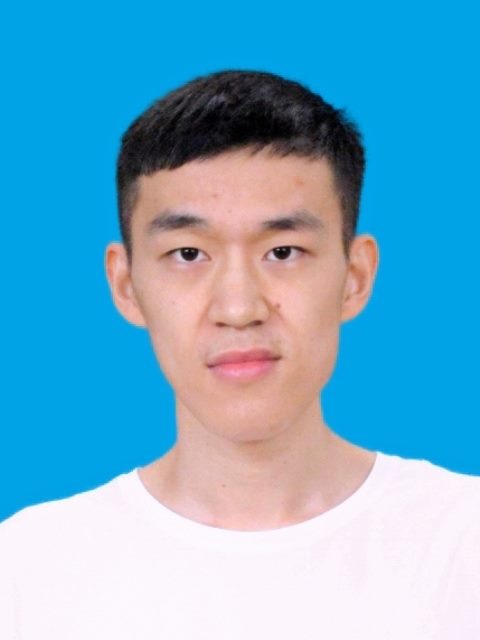}}]{Qi Hu} is pursuing his master degree in computer science at Huazhong University of Science and Technology (HUST), under the supervision of Dr. Bolong Zheng. He received his bachelor degree in Geographic Information Science (GIS) from Wuhan University in 2019. His research interests include spatio-temporal data management, fleet management, and reinforcement learning. 
\end{IEEEbiography}

\vspace{-42pt}

\begin{IEEEbiography}[{\includegraphics[width=1in,height=1.25in,clip,keepaspectratio]{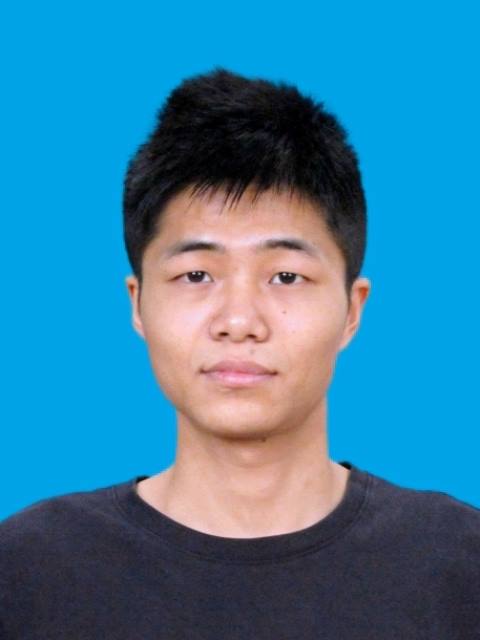}}]{Lingfeng Ming} is pursuing his master degree in computer science at Huazhong University of Science and Technology (HUST), under the supervision of Dr. Bolong Zheng. He received his bachelor degree in Energy and Power Engineering from Huazhong University of Science and Technology in 2019. His research interests include spatio-temporal data management, fleet management, and reinforcement learning. 
\end{IEEEbiography}

\vspace{-42pt}

\begin{IEEEbiography}[{\includegraphics[width=1in,height=1.25in,clip,keepaspectratio]{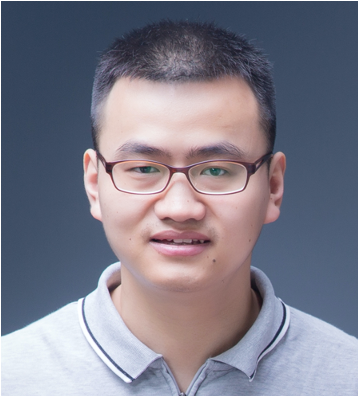}}]{Jilin Hu}
received the PhD degree from Aalborg University, Denmark, in 2019.
He is currently an assistant professor at Aalborg University, Denamrk. His research interests
include spatio-temporal data management, traffic data analytics, and machine learning.
He was a session chair for PVLDB'20. He has been reviewers for several top tier journals, e.g., IEEE TKDE, VLDB Journal, IEEE TNNLS, Neurocomputing, etc. He was also PC members for CVPR'21, AAAI'21, APWeb'20.
\end{IEEEbiography}

\vspace{-42pt}

\begin{IEEEbiography}[{\includegraphics[width=1in,height=1.25in,clip,keepaspectratio]{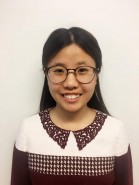}}]{Lu Chen}
received the PhD degree in computer science from Zhejiang University, China, in 2016.
She was an assistant professor in Aalborg University for a 2-year period from 2017 to 2019, and she was an associated professor in Aalborg University for a 1-year period from 2019 to 2020.
She is currently a ZJU Plan 100 Professor in the College of Computer Science, Zhejiang University, Hangzhou, China. Her research interests include indexing and querying metric spaces, graph databases, and database usability.
\end{IEEEbiography}

\vspace{-42pt}

\begin{IEEEbiography}[{\includegraphics[width=1in,height=1.25in,clip,keepaspectratio]{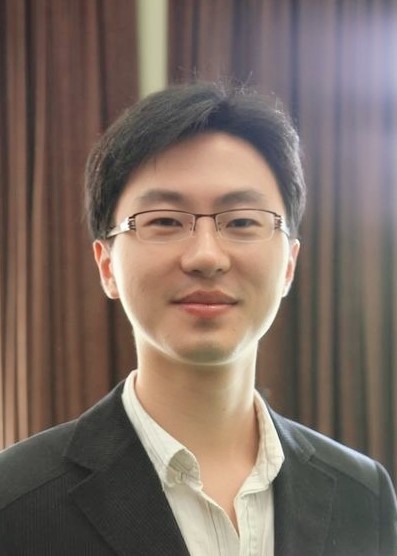}}]{Kai Zheng}
is a Professor of Computer Science with University of Electronic Science and Technology of China. He received his PhD degree in Computer Science from The University of Queensland in 2012. He has been working in the area of spatial-temporal databases, uncertain databases, social-media analysis, and blockchain technologies. He has published over 100 papers in prestigious journals and conferences in data management field such as SIGMOD, ICDE, VLDB Journal, ACM Transactions and IEEE Transactions. He is a senior member of IEEE.
\end{IEEEbiography}

\vspace{-42pt}

\begin{IEEEbiography}[{\includegraphics[width=1in,height=1.25in,clip,keepaspectratio]{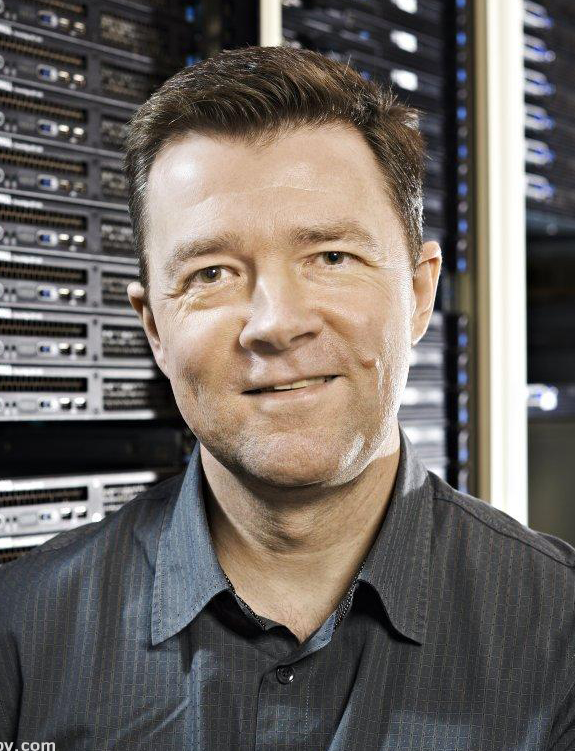}}]{Christian S. Jensen}
is a Professor of Computer Science at Aalborg University, Denmark. His research concerns data analytics and management with focus on temporal and spatio-temporal data management. Christian is an ACM and an IEEE fellow, and he is a member of the Academia Europaea, the Royal Danish Academy of Sciences and Letters, and the Danish Academy of Technical Sciences. He has received several national and international awards for his research, most recently the 2019 TCDE Impact Award. He was Editor-in-Chief of ACM TODS from 2014 to 2020 and an Editor-in-Chief of The VLDB Journal from 2008 to 2014.
\end{IEEEbiography}
\end{document}